\documentclass[11pt,a4paper]{article}

\usepackage[round]{natbib}

\usepackage{color,soul}

\usepackage{authblk}

\usepackage{graphics}
\usepackage{epsfig}
\usepackage{rotating}
\usepackage{amssymb}

\usepackage{supertabular}

\usepackage{longtable}

\usepackage{lscape}

\usepackage{fullpage}

\begin{document}




\title{Phase reddening on near-Earth asteroids: Implications for mineralogical analysis, space weathering and taxonomic classification}

\author[1,2]{Juan A. Sanchez\footnote{E-mail address: sanchez@mps.mpg.de}}
\author[3,1]{Vishnu Reddy}
\author[1]{Andreas Nathues}
\author[4]{Edward A. Cloutis}
\author[4]{Paul Mann}
\author[2]{Harald Hiesinger}

\affil[1]{Max Planck Institut f$\ddot{u}$r Sonnensystemforschung, Max Planck Str.2, 37191 Katlenburg-Lindau, Germany}

\affil[2]{Institut f$\ddot{u}$r Planetologie, 48149 M$\ddot{u}$nster, Germany}

\affil[3]{Department of Space Studies, University of North Dakota, Grand Forks, ND 58202, USA}

\affil[4]{Department of Geography, University of Winnipeg, Winnipeg, Manitoba, Canada}

\date{}

\maketitle

\begin{abstract}

 Phase reddening is an effect that produces an increase of the spectral slope and variations in the strength of the absorption bands as the phase angle increases. In order to understand  
 its effect on spectroscopic observations of asteroids, we have analyzed the visible and near-infrared spectra (0.45-2.5 $\mu$m) of 12 near-Earth asteroids observed at different phase angles. All these 
 asteroids are classified as either S-complex or Q-type asteroids. In addition, we have acquired laboratory spectra of three different types of ordinary chondrites at phase angles ranging from $13^\mathrm{o}$ to 
 $120^\mathrm{o}$. We have found that both, asteroid and meteorite spectra show an increase in band depths with increasing phase angle. In the case of the asteroids the Band I depth increases 
 in the range of $\sim$ $2^\mathrm{o}<$ g $< 70^\mathrm{o}$ and the Band II depth increases in the range of $\sim$ $2^\mathrm{o}<$ g $< 55^\mathrm{o}$. 
 Using this information we have derived equations that can be used to correct the effect of phase reddening in the band depths. 
 Of the three meteorite samples, the (olivine-rich) LL6 ordinary chondrite is the most affected by phase reddening. The studied ordinary chondrites have their maximum spectral contrast of Band I depths at a phase 
 angle of $\sim$ $60^\mathrm{o}$, followed by a decrease between $60^\mathrm{o}$ and $120^\mathrm{o}$ phase angle. The Band II depths of these samples have their 
 maximum spectral contrast at phase angles of  $30^\mathrm{o}-60^\mathrm{o}$ which then gradually decreases to $120^\mathrm{o}$ phase angle. The spectral slope of the ordinary chondrites spectra 
 shows a significant increase with increasing phase angle for g $> 30^\mathrm{o}$. Variations in band centers and band area ratio (BAR) values were also found, however they seems to have no significant impact on 
 the mineralogical analysis. Our study showed that the increase in spectral slope caused by phase reddening is comparable to certain degree of space weathering. In particular, an increase in phase angle in the range 
 of $30^\mathrm{o}$ to $120^\mathrm{o}$ will produce a reddening of the reflectance spectra equivalent to exposure times of $\sim 0.1\times 10^{6}$ to $1.3\times 10^{6}$ years at about 1 AU from the Sun. 
 This increase in spectral slope due to phase reddening is also comparable to the effects caused by the addition of different fractions of SMFe. Furthermore, we found that under some circumstances phase reddening 
 could lead to an ambiguous taxonomic classification of asteroids.  
  
\end{abstract}

\clearpage

\section{Introduction}

Near-infrared (NIR) spectroscopy is a powerful technique to derive information about the surface mineralogy of asteroids.  This mineralogical characterization relies  primarily on the identification 
and analysis of diagnostic features present in the spectra of some asteroids. Olivine and pyroxene are two common mafic minerals found on asteroids \citep{2002aste.conf..183G}, 
and their spectral properties dominate the reflectance spectra of some particular classes. Asteroids belonging to the S-complex and Q-types \citep[Bus-DeMeo taxonomy hereafter]{2009Icar..202..160D}
are examples of these objects. In the case of olivine, the absorption feature is composed of three overlapping bands and is centered near 1.04-1.1 $\mu$m, while pyroxenes show to major absorption bands, 
centered near 0.9-1 $\mu$m and 1.9-2 $\mu$m, all of them caused by the presence of $\rm{Fe^{2+}}$ cations \citep[e.g.,][]{1974JGR....79.4829A, 1975Adams, 1993macf.book.....B}. In the 
spectra of olivine-pyroxene mixtures the wavelength position (band centers) of the combined absorption features near 1 $\mu$m (Band I) is a function of relative abundance and composition of olivine and pyroxene, 
while the position of the feature near 2 $\mu$m (Band II) is a function of the pyroxene composition \citep[]{1986JGR....9111641C,1993Icar..106..573G}. In addition to the absorption band centers, the ratio of the area 
of Band II to that of Band I, known as band area ratio (BAR), is used to estimate olivine and pyroxene abundances in asteroids and meteorites  \citep{1986JGR....9111641C,2003AMR....16..185B,2010Icar..208..789D}. 
Together, these spectral band parameters constitute a useful tool to obtain information about the surface mineralogy and composition of asteroids.  However, band parameters can also be affected by other factors not 
related to compositional variations. One of these factors is the phase angle at the moment of the observation. The phase angle (g), is defined as the angular separation between the Sun and the observer as seen from the 
target. Phase angle induced effects can manifest themselves as phase reddening, which produces an artificial increase (reddening) of the spectral slope and variations in the strength of the absorption bands with increasing 
phase angle. This effect is explained as the result of the wavelength dependence of the single-scattering albedo \citep{1980LPSC...11..799G,1986Icar...66..455G,2002Icar..155..189C}. Traditionally, the term phase 
reddening was used to describe only the increase of the spectral slope or continuum, however as will be shown later, as the phase angle increases variations in the absorption bands are also seen. Therefore, in the present 
work the term phase reddening is extended to refer not only to the increase of the spectral slope, but also to the variations in the strength of the absorption bands. 

Phase reddening effect among Solar System objects was first noticed in asteroid broad band colors. \citet{1970sips.conf..317G} reported a phase reddening in the B-V and U-B colors of 4 Vesta to be 0.0018 and 
0.0027 mag/degree, respectively. Additional studies using photometric observations include the works of  \citet{1976Icar...28...53M} and \citet{1981AJ.....86.1694L}. The effects of phase reddening have been also 
observed among NEAs. \citet{1990AJ.....99.1985L} found that the spectral slopes of NEAs in general are higher than those measured for 3:1 resonance asteroids. Since NEAs are often observed at high phase angles 
this increase in the spectral slopes was interpreted as phase reddening.

\citet{Nathues00, 2010Icar..208..252N} carried out a spectroscopic and spectrophotometric survey of the Eunomia asteroid family, obtaining spectra in visible (VIS) and NIR wavelengths of 97 of its members. The 
analysis of the spectral slopes of these asteroids revealed an average increase of 0.067\%/100 nm per degree with increasing phase angle in the range of $2^\mathrm{o}<$ g $< 24^\mathrm{o}$. Apart from the increase 
of the spectral slope he also observed an increase of the 1 $\mu$m absorption band (depth) with increasing phase angle.  

\citet{2012Icar..217..153R} conducted an extensive study of Asteroid (4) Vesta in order to quantify phase angle-induced spectral effects on this asteroid prior to the arrival of the Dawn spacecraft. 
Rotationally resolved NIR spectral observations (0.7-2.5 $\mu$m) were obtained for this purpose. They found that in the phase angle range of $0^\mathrm{o}<$ g $ \le 25^\mathrm{o}$ for every $10^\mathrm{o}$ increase 
in phase angle, Vesta's  Band I and Band II depths increase 2.35\% and 1.5\%, respectively, and the BAR value increases by 0.30.

Phase reddening effect has been also detected and quantified from spacecraft observations of asteroids. The NEAR Shoemaker spacecraft obtained NIR spectroscopic observations (0.8-2.4 $\mu$m) 
of Asteroid (433) Eros at phase angles ranging from $0^\mathrm{o}$ to $100^\mathrm{o}$. These observations showed the most intense phase reddening for wavelengths inside of the 1.0 $\mu$m band, 
occurring at the level of 10\% across this phase angle range \citep{2002Icar..155..189C, 2002Icar..155..119B}. Similar results were observed from NIR reflectance spectra of Asteroid (25143) Itokawa acquired 
by the Hayabusa spacecraft \citep{2008Icar..194..137K}.

Laboratory measurements of different materials have confirmed the existence of the phase reddening effect.
\citet{1980LPSC...11..799G} investigated phase angle induced effects on the spectrophotometric properties of powdered materials. They obtained reflectance spectra at phase angles of 
$4^\mathrm{o}\le$ g $\le 120^\mathrm{o}$, in the wavelength range $0.4 \le \lambda \le 1.2$ $\mu$m. All the samples studied by them showed a significant reddening as the phase 
angle increases from 4 to $120^\mathrm{o}$. Furthermore, some of the spectra showed variations in the spectral contrast of the absorption bands with increasing phase angle. 
More recent laboratory studies related to phase reddening include the works of \citet{2011LPI....42.2268M} and \citet{2011LPI....42.1043S}.

Despite the fact that phase reddening has been known for a long time, its effect on the analysis of asteroid spectra has not been fully assessed. In the present work we study 
the phase reddening on NEAs and its implications for mineralogical analysis, space weathering and taxonomic classification. Due to their proximity, NEAs can exhibit a wide range of phase angles, 
often much higher than those observed for main belt asteroids, which make them the logical choice for this study. Our investigation focuses on the analysis of 27 VIS-NIR spectra of 12 NEAs previously 
classified as either S-complex or Q-types (Bus-DeMeo), which have been observed at different phase angles. In addition to the ground-based observations, laboratory spectra of ordinary 
chondrites are also analyzed and the results compared to those obtained from the asteroid spectral data.

\label{}

\section{Phase reddening from ground-based observations of NEAs}

\subsection{The data}

One of the drawbacks of this kind of study is the necessity to obtain multiple observations of the 
same asteroids at different phase angles, which in terms of allocated observation time can be very difficult to do. For this reason we have decided to combine data from 
\citet{Reddy09}, \citet{2009Icar..202..160D}, \citet{2007M&PS...42.2165A}, \cite{2001M&PS...36.1167B} and the MIT-UH-IRTF Joint Campaign for NEO Spectral Reconnaissance (NEOSR).
All of these observations were conducted with the SpeX instrument \citep{2003PASP..115..362R} on the NASA Infrared Telescope Facility (IRTF).
The NIR spectra (0.7-2.5 $\mu$m) were obtained using Spex in its low resolution ($R$ $\sim$ 150) prism mode with a 0.8" slit width. 
Typical observations consist of acquiring frames employing a nodding sequence in which the object is alternated between two different slit positions (A and B) following the sequence ABBA.
Since the object and the sky are observed simultaneously in each exposure, by subtracting A from B and B from A, is possible to remove the sky background during the data reduction process.
In order to minimize the effects of differential atmospheric refraction the slit is oriented along the parallactic angle during the observations.
To correct for telluric water vapor features, and to obtain the relative reflectance, local standard and solar analog stars are observed at airmasses as similar as possible 
as the asteroids. Flat fields and arc line spectra for each night are also obtained during the observations. The data reduction is performed 
using different packages, including IRAF and Spextool  \citep{2004PASP..116..362C}. For more details about the observing protocols and reduction 
of these data sets  see \citet{Reddy09}, \citet{2009Icar..202..160D}, \citet{2007M&PS...42.2165A} and \cite{2001M&PS...36.1167B}.     
The spectra of the NEAs at visible wavelengths used in the present work were taken from \cite{2001M&PS...36.1167B, 2004Icar..170..259B} and \citet{2009Icar..202..160D}. All the 
spectra were normalized to unity at 0.55 $\mu$m. Observational circumstances for the studied objects are presented in Table \ref{t:Table1}.

\subsection{Spectral band analysis of NEAs}

As mentioned earlier, increasing the phase angle will produce an increase in the spectral slope and changes in the strength of the absorption 
bands. Spectral slope variations can be interpreted as surface heterogeneities caused by different factors such as metal content, particle 
size and space weathering \citep{1993Icar..106..573G, 2002aste.conf..585C, 2010Icar..209..564G}.
However, this parameter is also known to be very sensitive to other factors. Apart from the viewing geometry,  atmospheric differential refraction 
\citep{1982PASP...94..715F}, airmass differences between the standard star and the asteroid at the time of the observations \citep{2011Icar..212..677D}, incorrect 
centering of the object in the slit \citep{2004PASP..116..362C}, poor weather conditions \citep{2002aste.conf..169B} and the use of different solar analogs 
\citep{2004M&PS...39.1343S} are often the cause of fluctuations in the asteroid's spectral slope. In some cases the origin of these error sources is 
difficult to determine since they are seen as nonsystematic variations in the final asteroid reflectance spectrum \citep{2006Icar..181...94H}. Therefore, in order to better 
quantify the effect of phase reddening on the spectral slope we will leave that part of the study to the laboratory spectra (see section 3), and will focus our attention on 
the analysis of the absorption bands of the asteroid spectra.

The spectral band parameters for each VIS-NIR spectrum were measured using the Spectral Processing Routine (SpecPR) program based on the protocols discussed by 
\citet{1986JGR....9111641C}, \citet{2002aste.conf..183G}, and \citet{2003LPI....34.1602G}. Band centers are calculated by dividing out the 
linear continuum and fitting an n-order polynomial over the bottom third of each band. In the case of Band II, we defined the unresolved red edge as 2.44 $\mu$m.
After ratioing the absorption features to the straight-line continuum, a subroutine in SpecPR was used to calculate the band areas (areas between the linear continuum and the data curve) 
and band depths (measured from the continuum to the band center and given as percentage depths). The band areas were then used to calculate the BAR values.
Each band parameter was measured ten times using different order polynomial fits (typically third and fourth order) and sampling different ranges of points 
within the corresponding intervals. Averages of these measurements were taken as the final values.  The uncertainties are given 
by the average 1-$\sigma$ (standard deviation of the mean) calculated from the multiple measurements of each band parameter. 
The olivine-pyroxene abundance ratio of the asteroids was estimated using the relationship between $ol/(ol+px)$ and BAR derived by \citet{2010Icar..208..789D} from 
the analysis of 48 ordinary chondrite samples. This linear regression is expressed as

\begin{equation}
ol/(ol+px) = -0.242\times BAR+0.728
\end{equation}

where the $ol/(ol+px)$ is expressed as a decimal. The band parameters and $ol/(ol+px)$ ratios with their corresponding errors obtained for each asteroid at different phase angles are 
presented in Tables \ref{t:Table2} and \ref{t:Table3} respectively. 

\subsection{Temperature correction to band parameters}

Temperature-induced effects on the spectra of mafic minerals are characterized by shifting the band centers and broadening or narrowing the 
absorption features \citep{1985JGR....9012434S,1999AdSpR..23.1253S,2000Icar..147...79M,2002Icar..155..169H}. 
While surface temperature variations are considered of minor importance for main-belt asteroids, they may be significant for objects with higher 
eccentric orbits like NEAs \citep{2000Icar..147...79M,2000LPI....31.1003S}. 
Furthermore, the spectra of laboratory samples are commonly obtained at room temperature ($\sim$ 300 K). Thus, in order to compare the band 
parameters with those measured for laboratory samples temperature corrections should be applied.

For each asteroid we have determined the average surface temperature ($T$) in the same way as \citep[e.g.,][]{2009M&PS...44.1331B}, where the temperature of the asteroid is approximated 
by the equation for energy conservation. The calculated temperatures of the NEAs are given in Table \ref{t:Table2}.

Since all the asteroids studied in this work are S-complex or Q-types, i.e., olivine-orthopyroxene assemblages, we have derived temperature corrections based on 
the analysis of ordinary chondrites. These meteorites are the most common type of meteorite to fall on Earth ($\sim$ 85\% of all), and they have been generally linked 
to either S-complex or Q-type asteroids \citep[e.g.,][]{1993Icar..106..573G, 2008Natur.454..858V, 2011Sci...333.1113N}. 

We have reanalyzed spectra of two H5 ordinary chondrites from \citet {2002Icar..155..169H} acquired in the temperature range between 80 and 400 K. Spectral band 
parameters (band centers, band depths and BAR) and their uncertainties were measured using the same methods applied to the asteroid spectra. These data were 
combined with the results obtained by  \citet{2000Icar..147...79M} for a L5 and a LL4 ordinary chondrite acquired at temperatures between 293 and 80 K.

We found that Band II is the most affected by temperature variations. A correlation between Band II centers and temperature is observed for all samples, i.e., Band II centers shift 
to longer wavelengths as the temperature increases. Therefore, to each data set a linear fit was performed and the equations that represent those linear fits were averaged, obtaining 
the following expression: 

\begin{equation}
BII(T) = 0.0002\times T(K)+1.87
\end{equation}

where the Band II center (BII) is given in $\mu$m. From this equation we derived a wavelength correction for the Band II center

\begin{equation}
\Delta BII(\mu m) = 0.06-0.0002\times T(K)
\end{equation}

This correction is derived with respect to room temperature (300 K) and must be added to the calculated Band II center of each asteroid. The temperature corrected Band II centers are 
presented in Table \ref{t:Table2}. For the Band I center we found that the wavelength shift is in general very small ($\sim$ 0.003 $\mu$m), so no temperature correction was derived.
 
A decrease in Band II depth with increasing temperature was found for the three groups of ordinary chondrites. No obvious trend was observed for the Band I depth. 
Applying a similar procedure to the Band II center, we can estimate the approximate rate of change of the Band II depth for a range of temperatures 

\begin{equation}
BII_{dep}(T) = 20.89-0.015\times T(K)
\end{equation}

where the Band II depth (BII$_{\rm{dep}}$) is given in \%. From this equation we derived a temperature correction for the Band II depth

\begin{equation}
\Delta BII_{dep}(\%) = 0.015\times T(K)-4.5
\end{equation}

This correction is derived with respect to the room temperature and must be added to the calculated Band II depth of each asteroid.

An inverse correlation between BAR and temperature was also found for the ordinary chondrites samples.
As in the previous cases we obtained an average expression to calculate the rate of change of BAR for a range of temperatures

\begin{equation}
BAR(T) = 0.83-0.00075\times T(K)
\end{equation}

from which a temperature correction with respect to room temperature was derived

\begin{equation}
\Delta BAR = 0.00075\times T(K)-0.23
\end{equation}

This correction must be added to the calculated BAR values of each asteroid before comparing with those calculated for laboratory samples obtained at room temperature.
The temperature corrected Band II depths and BARs are presented in Table \ref{t:Table2}. Based on the corrected BAR values, the olivine abundance of 
each asteroid was recalculated using Equation 1. These values are included in Table \ref{t:Table3}.

\begin{figure*}[!ht]
\begin{center}
\psfig{file=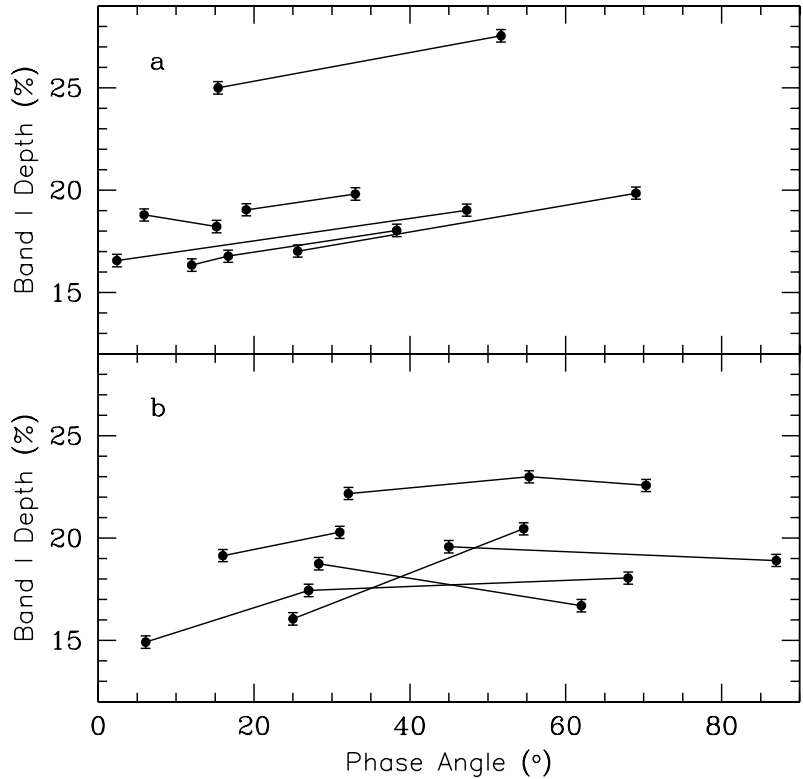,angle=0,height=11cm}
\caption{\label{f:DBIphase6} {\small Band I depth as a function of phase angle for the NEAs. Values derived from multiple observations of individual asteroids are connected by lines and have been displayed 
into two panels for clarity.}}
\end{center}
\end{figure*}

\subsection{Phase reddening effect on the band parameters}

\begin{figure*}[!ht]
\begin{center}
\psfig{file=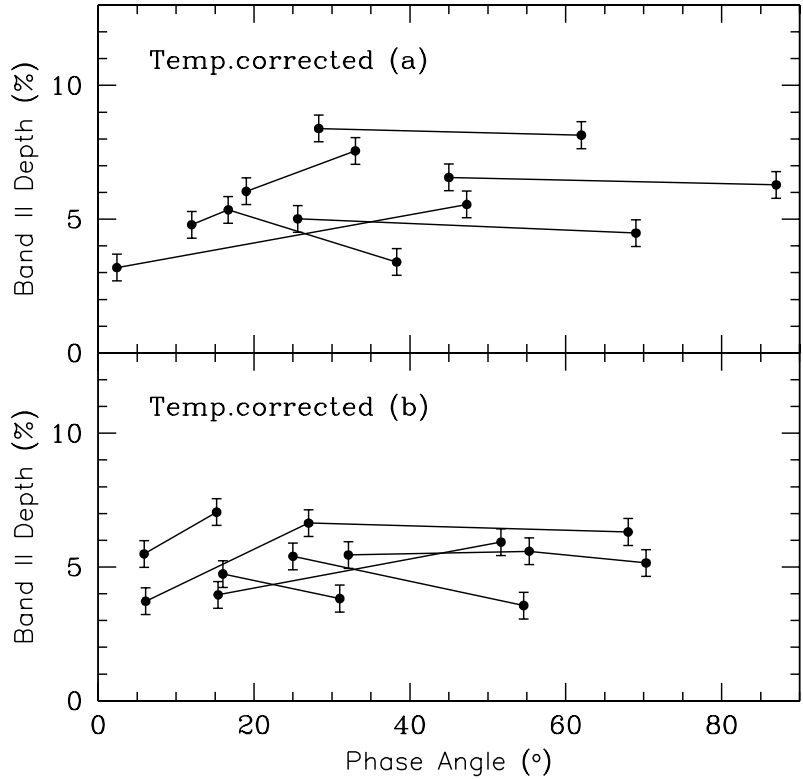,angle=0,height=11cm}
\caption{\label{f:DBIIphase9} {\small Temperature corrected Band II depth as a function of phase angle for the NEAs. Values derived from multiple observations of individual asteroids are connected by lines and have 
been displayed into two panels for clarity.}}
\end{center}
\end{figure*}

An inspection of the measured Band I depths presented in Table \ref{t:Table2} indicates a change of this parameter with phase angle. These values are plotted as a function 
of phase angle in Figure \ref{f:DBIphase6}. The black circles represent the measured Band I depths. Values derived from multiple observations of individual asteroids are 
connected by lines and have been displayed into two panels for clarity. A tendency of increasing spectral contrast as the phase angle increases is observed for most of the objects
in the range of $\sim$ $2^\mathrm{o}<{\rm{g}}< 70^\mathrm{o}$. When higher phase angles are included (up to $\sim$ 90$^\mathrm{o}$) a slight decrease in Band I depth is observed.  
However, as we do not have sufficient data at such high phase angles, it remains unclear whether this is the actual tendency for ${\rm{g}}>70^\mathrm{o}$. 
A similar situation is observed with the temperature corrected Band II depths (Fig. \ref{f:DBIIphase9}). In this case the absorption band seems to increase in the range of 
$\sim$ $2^\mathrm{o}<{\rm{g}}< 55^\mathrm{o}$ and then it remains more or less constant for higher phase angles. In some cases we notice a deviation from these trends among the measured band depths, i.e.,   
a decrease of band depths with increasing phase angle. Although compositional variations cannot be completely ruled out, this is more likely due to the fact that some 
spectra show more scattering than others, especially in the Band II. This is probably caused by an incomplete correction of the telluric water bands, and due to the decreased response of the detector 
for wavelengths beyond 2.4 $\mu$m. This was seen in the Band I of Asteroids 1036 Ganymed (${\rm{g}}=15.2^\mathrm{o}$) and 4954 Eric (${\rm{g}}=62^\mathrm{o}$) and in the Band II of Asteroids 
1620 Geographos (${\rm{g}}=38.3^\mathrm{o}$), 1627 Ivar (${\rm{g}}=31.0^\mathrm{o}$) and 1980 Tezcatlipoca (${\rm{g}}=54.6^\mathrm{o}$).

\begin{figure*}[!ht]
\begin{center}
\psfig{file=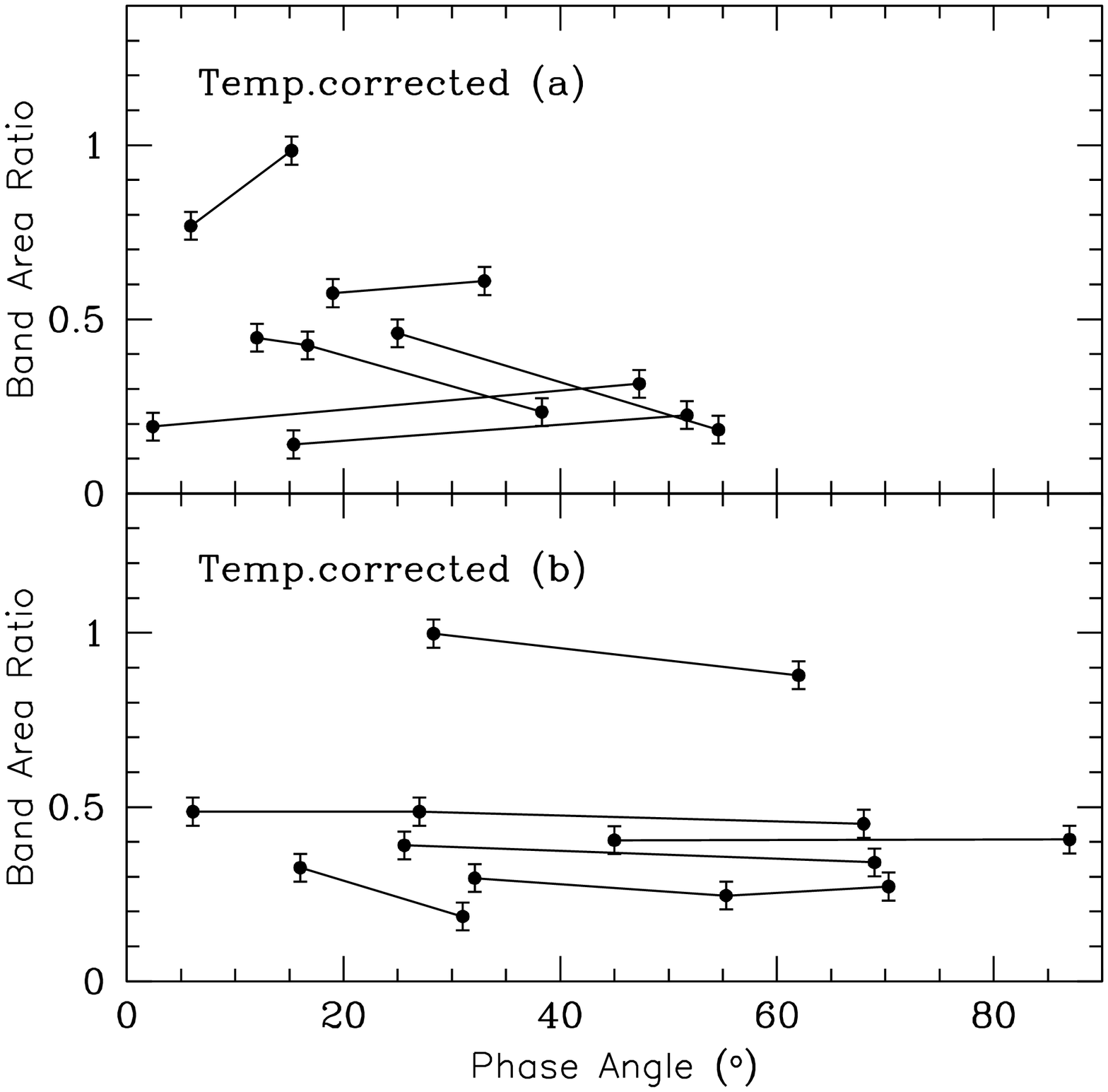,angle=0,height=11cm}
\caption{\label{f:BARphase6} {\small Temperature corrected BAR values as a function of phase angle for the NEAs. Values derived from multiple observations of individual asteroids are connected by lines and have 
been displayed into two panels for clarity.}}
\end{center}
\end{figure*}

\begin{figure*}[!ht]
\begin{center}
\psfig{file=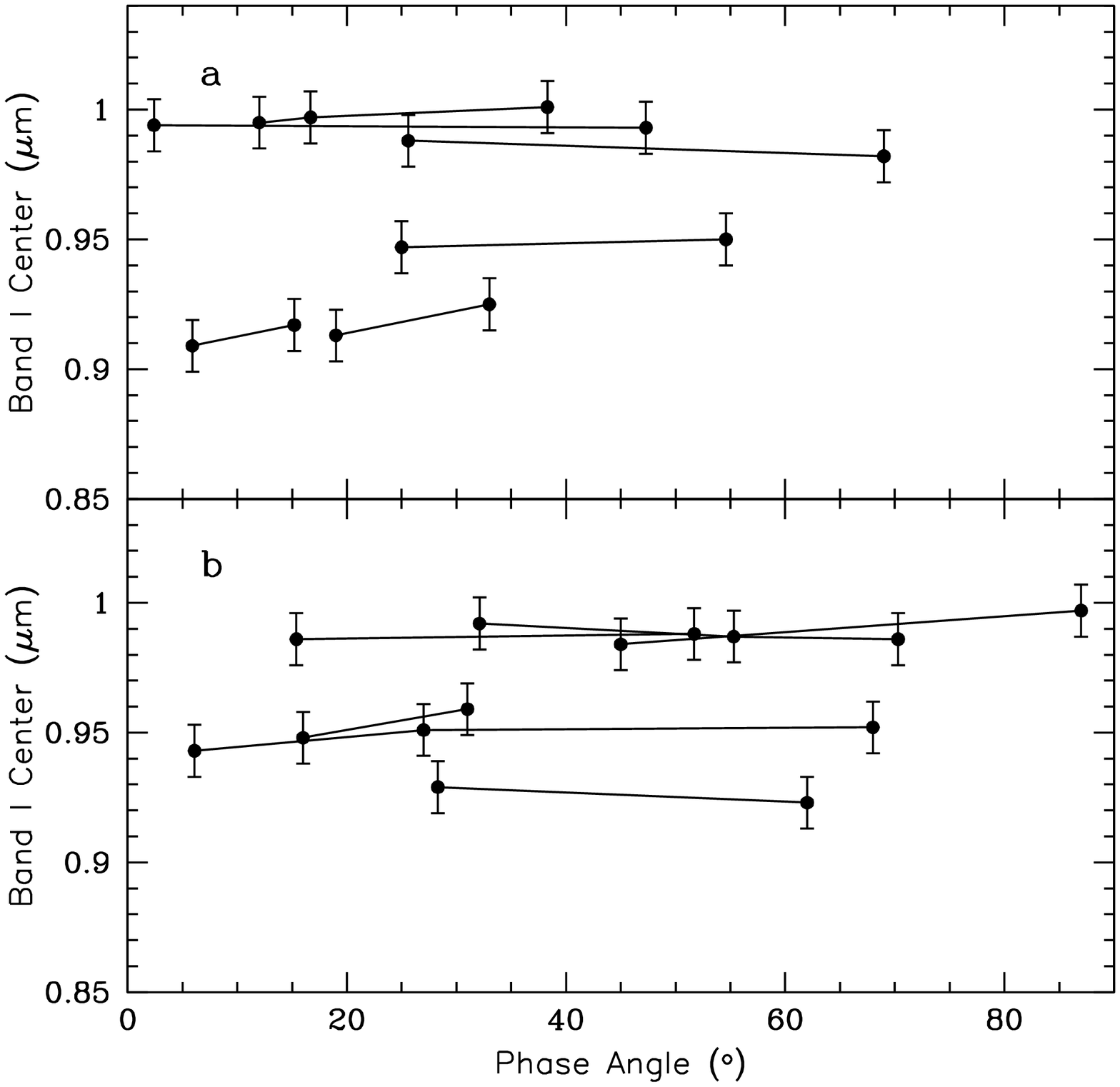,angle=0,height=11cm}
\caption{\label{f:BIphase5} {\small Band I center as a function of phase angle for the NEAs. Values derived from multiple observations of individual asteroids are connected by lines and have been displayed into two 
panels for clarity.}}
\end{center}
\end{figure*}

In an effort to quantify the phase reddening effect on band depths we performed a linear fit to the data of each asteroid. Only the data in the phase angle ranges where the correlations are observed were fitted. 
Those values that deviate from the general tendency were not consider. The equations derived from the linear fits were then averaged to obtain a general expression for each band depth. 
Previous works have derived correlations between band parameters and phase angle \citep[e.g.,][]{1990AJ.....99.1985L,2010Icar..208..773M}. However, these correlations were obtained from mixed observations of 
different objects, making it very difficult to disentangle the phase reddening from other effects. Our approach, on the other hand, attempts to overcome this problem by evaluating the phase reddening on individual 
objects from which an average expression is then derived. We caution that this procedure only provides a rough estimation of the effects of phase reddening on band depths, since for each asteroid we have a limited phase 
angle range. The general equations for the band depths are given by

\begin{equation}
BI_{dep} ({\rm{g}}) = 0.066\times({\rm{g}})+17.42
\end{equation}

\begin{equation}
BII_{dep} ({\rm{g}}) = 0.093\times({\rm{g}})+3.73
\end{equation}

where the band depths are measured in \%. Thus, according to these equations, on average, Band I and Band II depths will increase 0.66\% and 0.93\% respectively, for every 10$^\mathrm{o}$ 
increase in phase angle in the range of $2^\mathrm{o}<{\rm{g}}< 70^\mathrm{o}$ for Band I and  $2^\mathrm{o}<{\rm{g}}< 55^\mathrm{o}$ for Band II.  

From equations (8) and (9) we can obtain expressions for band depth corrections that are a function of the phase angle,

 \begin{equation}
BI_{depc} = BI_{dep}-0.066\times({\rm{g}})
\end{equation}

 \begin{equation}
BII_{depc} = BII_{dep}-0.093\times({\rm{g}})
\end{equation}

where BI$_{\rm{depc}}$ and BII$_{\rm{depc}}$ are the corrected band depths. With these equations we can roughly correct the effect of phase reddening in the band depths in the phase angle ranges of 
$2^\mathrm{o}<{\rm{g}}< 70^\mathrm{o}$ for Band I and  $2^\mathrm{o}<{\rm{g}}< 55^\mathrm{o}$ for Band II. 
 
The temperature corrected BAR values presented in Table \ref{t:Table2} and plotted in Figure \ref{f:BARphase6} indicate variations with increasing phase angle, however no obvious trend can be seen.
The most significant change in BAR is observed for the same objects whose absorption bands show more scattering. 

Small variations on band centers with increasing phase angle were found, however they are within the uncertainties associated to these band parameters. This can be seen in Figure \ref{f:BIphase5} 
where we plotted Band I center as a function of phase angle for the NEAs.  The black circles represent the measured Band I centers. Values derived from multiple observations of the same asteroids are 
connected by lines.


\section{Phase reddening from laboratory measurements of ordinary chondrites}

\subsection{Data and spectral band analysis}

As it was stated before, ordinary chondrites are considered to be the meteorite analogs of S-complex and Q-type asteroids. Therefore, in order to complement our study we have analyzed the spectra of a group of 
ordinary chondrites obtained at a wide range of phase angles.   

 Diffuse reflectance spectra were collected at the University of Winnipeg Planetary Spectrophotometer Facility (UWPSF) using an ASD FieldSpec Pro HR spectrometer over the wavelength range of 0.35 to 2.5 $\mu$m.
 The three samples that have been analyzed are Dhurmsala (LL6, fell 1860), Pavlograd (L6, fell 1826), and Lancon (H6, fell 1897).  
 They were all crushed and sieved to a grain size of $<150$ $\mu$m. The samples were gently poured into aluminum sample cups and the edge of a glass slide was drawn  across the sample to provide a flat surface for 
 the spectral measurements.
 Reflectance spectra were acquired relative to a 100\% Labsphere Spectralon disk measured at an incident angle $i$=$13^\mathrm{o}$ and emission angle 
 $e$=$0^\mathrm{o}$ ($13^\mathrm{o}$ phase angle). Ten sets of measurements were acquired for each sample resulting in three emission angles ($e$=$0^\mathrm{o}$, 30$^\mathrm{o}$, 60$^\mathrm{o}$), 
 five incidence angles ($i$=0$^\mathrm{o}$, 13$^\mathrm{o}$, 30$^\mathrm{o}$,-30$^\mathrm{o}$, 60$^\mathrm{o}$) and five different phase angles ranging from 13$^\mathrm{o}$ to 120$^\mathrm{o}$. 
 Positive angles ($i$ and $e$) are measured when the light source and the detector are on either side of the normal, while negative incidence angles are measured when both light source and detector are on the same side 
 of the normal. For each measurement, a total of 250 scans were collected and averaged to improve the signal to noise ratio. 
 
Spectral band parameters and their uncertainties were measured for each VIS-NIR spectrum using the same methods applied to the asteroid spectra. In addition to the band parameters we have also measured the 
spectral slope, which was determined from the fitted continuum across Band I, i.e., a straight line tangent to the reflectance peaks from $\sim$ 0.7 to $\sim$ 1.55 $\mu$m. The uncertainty of the spectral 
slope is given by the average 1-$\sigma$, estimated from sampling different ranges of points near to the reflectance peaks on either side of the absorption band. The olivine-pyroxene abundance ratio of the samples 
was estimated using Eq.(1). The band parameters, spectral slopes and $ol/(ol+px)$ ratios with their corresponding errors obtained for each sample are presented in Table \ref{t:Table4}.  

\subsection{Phase reddening effect on the band parameters}

The analysis of the laboratory spectra shows that variations in the band parameters can arise not only by changing the phase angle, but also for the same phase angle using different configurations of the 
incidence and emission angle. As can be seen in Table \ref{t:Table4} phase angles of 30$^\mathrm{o}$, 60$^\mathrm{o}$ and 90$^\mathrm{o}$ were obtained using different combinations of $i$ and $e$, and 
these different combinations produced, in some cases, very different band parameter values. Since we want to quantify the effects of phase reddening on the spectral band parameters, for those phase angles where 
more than one combination of $i$ and $e$ was used the average value for each band parameter was taken. These average values are presented in Table \ref{t:Table4}.

\begin{figure*}[!ht]
\begin{center}
\psfig{file=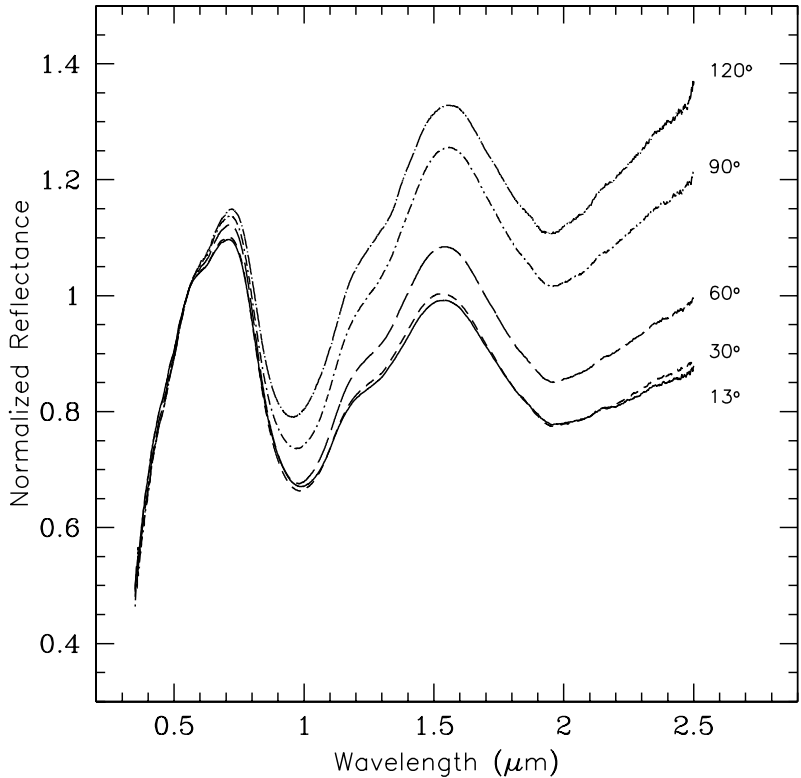,angle=0,height=11cm}
\caption{\label{f:LL6spectra3} {\small Reflectance spectra of the LL6 chondrite Dhurmsala obtained at five different phase angles. From the bottom to the top g=$13^\mathrm{o}$ (solid line),
g=$30^\mathrm{o}$ (short dashed line), g=$60^\mathrm{o}$ (long dashed line), g=$90^\mathrm{o}$ (dot-short dashed line) and g=$120^\mathrm{o}$ (dot-long dashed line). 
All the spectra are normalized to unity at 0.55 $\mu$m.}}
\end{center}
\end{figure*}

Figure \ref{f:LL6spectra3} shows the reflectance spectra of the LL6 ordinary chondrite at five different phase angles. From the bottom to the top the phase angles are 
$13^\mathrm{o}$, $30^\mathrm{o}$, $60^\mathrm{o}$, $90^\mathrm{o}$ and $120^\mathrm{o}$. The spectra corresponding to g = $30^\mathrm{o}$, $60^\mathrm{o}$ and $90^\mathrm{o}$ are the average 
spectra obtained from different combinations of the incidence and emission angles. All the spectra are normalized to unity at 0.55 $\mu$m. An increase in the spectral slope with increasing phase angle is 
evident, being more significant for phase angles higher than $30^\mathrm{o}$. A similar behavior was observed for the other two samples. For phase angles between $13^\mathrm{o}$ and $30^\mathrm{o}$ a slight 
reddening is observed in the LL6 spectra, while a slight blueing (i.e decreasing of the spectral slope with increasing phase angle) is seen in the L6 and H6 spectra. Variations in the strengths of the absorption bands 
are also seen for the three samples.

\begin{figure*}[!ht]
\begin{center}
\psfig{file=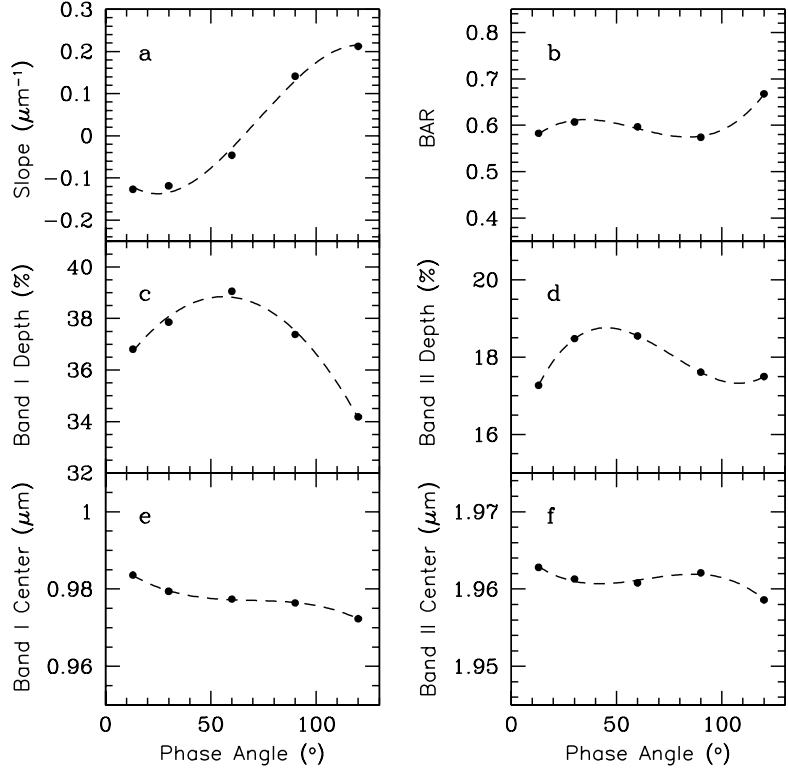,angle=0,height=11cm}
\caption{\label{f:LL6all} {\small LL6 ordinary chondrite: Measured spectral slopes (a), BAR values (b), Band I depths (c), Band II depths (d) Band I centers (e) and Band II centers (f) as a function of the phase angle. 
The error bars are smaller than the symbols. The dashed lines represent polynomial fits.}}
\end{center}
\end{figure*}

In Figure \ref{f:LL6all} we plotted the spectral slopes (panel a), the BAR values (panel b), the Band I depths (panel c), the Band II depths (panel d) and the band centers (panels e and f) as 
functions of the phase angle for the LL6 ordinary chondrite. The dashed lines are polynomial fits. 

From the "a" panel we can see that the measured spectral slopes remain more or less constant for phase angles between $13^\mathrm{o}$ and $30^\mathrm{o}$, and then they increase as the 
phase angle increases up to $120^\mathrm{o}$. A similar trend was observed for the L6 and H6 ordinary chondrites. The largest difference in spectral slope between the lowest ($13^\mathrm{o}$) and 
the highest ($120^\mathrm{o}$) phase angle was found for the LL6 ordinary chondrite, which increased 0.34 $\mu$$m^{-1}$, followed by the L6 ordinary chondrite with an increase of 
0.29 $\mu$$m^{-1}$ and the H6 ordinary chondrite with an increase of 0.16 $\mu$$m^{-1}$.
 
The Band I depth of the three samples has its maximum spectral contrast near g = $60^\mathrm{o}$. The LL6 spectra show a progressive increase of Band I depth from  g = $13^\mathrm{o}$ 
to $60^\mathrm{o}$ (Fig. \ref{f:LL6all} panel c), however the L6 and H6 spectra show almost no change in Band I depth for phase angles between $13^\mathrm{o}$ and $30^\mathrm{o}$, and then an increase 
in Band I depth from  g = $30^\mathrm{o}$ to $60^\mathrm{o}$. The spectra of the three samples show a decrease in Band I depth between $60^\mathrm{o}$ and $120^\mathrm{o}$ phase angle.  
 The largest increase in Band I depth between the lowest phase angle ($13^\mathrm{o}$) and the phase angle at which the Band I reaches its maximum spectral contrast ($\sim60^\mathrm{o}$ ) corresponds to the LL6 
 sample, with an increase of 2.25\%, followed by the L6 and H6 with an increase of 1.36 and 0.53\% respectively. Of the three samples, the LL6 also shows the most significant decrease of Band I depth between the phase 
 angle corresponding to the maximum spectral contrast and the highest phase angle ($120^\mathrm{o}$), with a decrease of 4.88\%, followed by the H6 with a decrease of 4.21\% and the L6 with a decrease of 3.89\%.
 
The behavior of the Band II depths seems to be more complex. The LL6 spectra show a progressive increase of Band II depths from  g = $13^\mathrm{o}$ to $\sim45^\mathrm{o}$ (where the band 
reaches its maximum spectral contrast) and then it drops between $45^\mathrm{o}$ and $90^\mathrm{o}$, becoming more or less constant from $90^\mathrm{o}$ to $120^\mathrm{o}$ phase angle 
(Fig. \ref{f:LL6all} panel d). 
The Band II depths of the L6 spectra show almost no change between $13^\mathrm{o}$ and $30^\mathrm{o}$. From $30^\mathrm{o}$ to $60^\mathrm{o}$ Band II depths increase and beyond $60^\mathrm{o}$ 
gradually decrease with phase angle increasing to $120^\mathrm{o}$. The Band II depths of the H6 spectra show a slight increase from $13^\mathrm{o}$ to $30^\mathrm{o}$ and then drop from $30^\mathrm{o}$ to 
$120^\mathrm{o}$. The most significant increase in Band II depths between the lowest phase angle and the phase angle at which the Band II reaches its maximum spectral contrast was observed for the LL6 sample, 
which shows an increase of 1.56\%, followed by the L6 with an increase in Band II depth of 0.56\% and the H6 with the lowest increase of 0.16\%. The largest difference in Band II depth between the phase angle 
corresponding to the maximum spectral contrast and the largest phase angle ($120^\mathrm{o}$) was observed for the L6 sample that shows a decrease of 3.11\%, followed by the H6 with a decrease of 3.02\% and 
the L6 with a decrease of 1.33\%.

Variations in the BAR values are observed among the three samples, being the maximum difference between the lowest and highest phase angle $\sim$0.1, however no systematic trends were found. A similar 
situation occur with the band centers, where the maximum shift (to shorter wavelengths) between the lowest and highest phase angle is $\sim$0.01 $\mu$m for the Band I center of the LL6 sample (Fig. \ref{f:LL6all} panel e).

\section{Phase reddening and mineralogical analysis}

The analysis of the laboratory samples indicate that the (olivine-rich) LL6 ordinary chondrite is the most affected by phase reddening. This sample shows the largest variation in spectral slopes and band 
depths with increasing phase angles. This material-dependence of the phase reddening confirm the results reported by previous studies \citep[e.g.,][]{1967JGR....72.5705A, 1980LPSC...11..799G}.
A comparison between the measurements obtained from the laboratory samples with those from the NEAs spectra, reveal that the rate of change in band depths for the ordinary chondrites is lower than that found for 
the asteroids. Differences in grain size, composition and packing between the NEAs studied and the meteorite samples could explain this discrepancy. Of the three samples, the spectral behavior of the LL6 is the closest 
to the NEAs. This resemblance could be attributed to the fact that half of the asteroids studied are located in the same region as the LL ordinary chondrites in the Band I center versus BAR diagram 
(see Figs. \ref{f:BI_BAR6B} and \ref{f:b1_BAROC3}). 

Because band centers and composition of mafic minerals are related, these band parameters are used to derive information about the pyroxene and olivine composition of asteroids surfaces.    
For that purpose, and from the analysis of laboratory samples, different empirical equations have been derived \citep[e.g.,][]{2002aste.conf..183G,2007LPI....38.2117B,2010Icar..208..789D}. 
Our results show that there is no significant change in band centers with increasing phase angle for both asteroid and meteorite spectra, being the largest shift of $\sim$ 0.01 $\mu$m (for the meteorite spectra), which is on 
the order of the uncertainty associated with this parameter for the asteroid spectra. Based on this, it seems to be unlikely that phase reddening could lead to a misinterpretation of the minerals composition in asteroids. 

In addition to the band centers, we used the BAR values and Eq. (1) to estimate olivine and pyroxene abundances for both NEAs and the laboratory samples. The results obtained (Tables \ref{t:Table3} and \ref{t:Table4}) 
show that the largest variation in the $ol/(ol+px)$ ratio with increasing phase angles for NEAs is $\sim$ 0.02 (not taken into account the values obtained from the noisy spectra), and for the 
ordinary chondrites between 0.03 and 0.05. Since these variations are on the order of the errors, effects of phase reddening on the estimation of the olivine-pyroxene abundance ratio seems to be negligible.

\section{Phase reddening and space weathering}

Space weathering is the term commonly used to refer to any process that modifies the surfaces of airless bodies. The effects of space weathering on the spectra can be seen as reddening of the spectral 
slopes and suppression of the absorption bands \citep{,2000M&PS...35.1101P,2001JGR...10610039H,2010Icar..209..564G}. 
The analysis of returned samples from the Moon and Asteroid (25143) Itokawa have shown that the cause of the spectral changes is the presence of submicroscopic metallic iron (SMFe) 
incorporated into the soil grains \citep{2000M&PS...35.1101P,2001M&PS...36..285T,2011Sci...333.1121N}. This SMFe is produced by condensation of vapors created by micrometeorite impacts and/or deposition 
of atoms sputtered off from silicates by solar wind ions  \citep{2001M&PS...36..285T,2001JGR...10610039H,2002aste.conf..585C}. Ion irradiation experiments conducted by \citet{2005Icar..179..265B} showed that solar 
wind irradiation can also redden reflectance spectra by creation of displacements (the sum of the vacancies and the replacements) caused by elastic collisions between ions and 
target nuclei. 

\citet{2006Icar..184..327B} determined that the effects of space weathering due to ion irradiation can be described by an exponential continuum. They computed the ratio between the reflectance spectra of irradiated and 
unirradiated samples and then modeled it with an exponential curve given by

\begin{equation}
{\rm{Ratio}} =  W(\lambda) = Kexp(C_{S}/\lambda)
\end{equation}

where $\lambda$ is the wavelength, K is a scale factor and the parameter C$_{S}$ is a measure of the effects of space weathering. They called $W$($\lambda$) the weathering function.
Since both effects, phase reddening and space weathering, are manifested in a similar way, we investigated whether the red slopes exhibited by spectra obtained at high phase angles could be misinterpreted as 
space weathering. In order to do this we carried out two different experiments. 

\begin{figure*}[!ht]
\begin{center}
\psfig{file=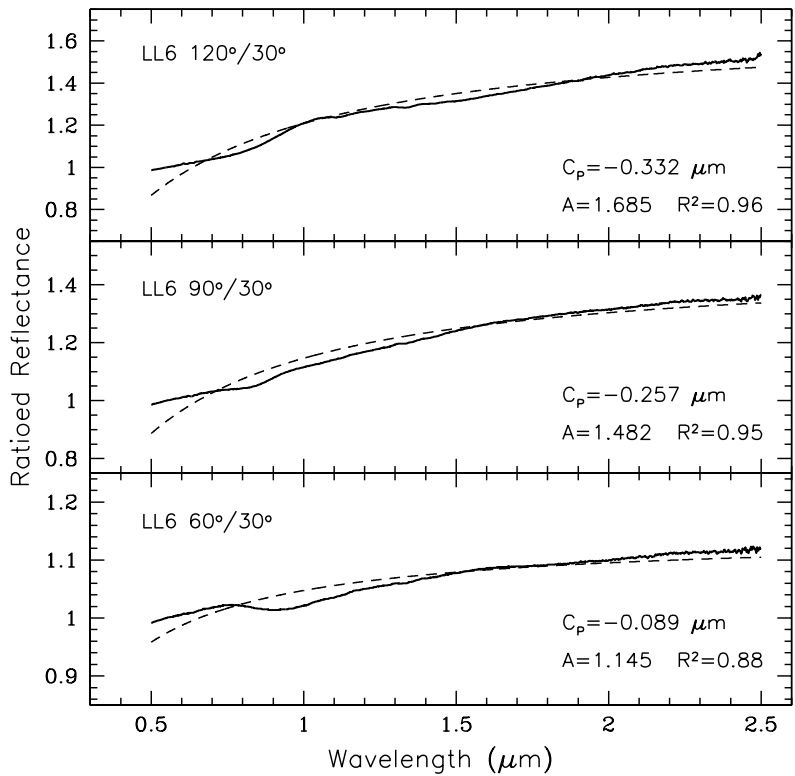,angle=0,height=11cm}
\caption{\label{f:RatioedR2} {\small Ratio plots (solid lines) between the spectra of the LL6 obtained at g=$60^\mathrm{o}$ and g=$30^\mathrm{o}$ (bottom), at g=$90^\mathrm{o}$ and 
g=$30^\mathrm{o}$ (middle) and at g=$120^\mathrm{o}$ and g=$30^\mathrm{o}$ (top panel). Dashed curves are the best fit curves obtained using Eq. (13). For each ratio plot the resultant 
A, C$_{P}$ and $R^{2}$ values are given.}}
\end{center}
\end{figure*}

For the first experiment we followed the same procedure used by \citet{2006Icar..184..327B}, with the difference that instead of dividing the  
reflectance spectra of an irradiated by an unirradiated sample, we computed the ratio between the spectra obtained at high phase angles and those obtained at low phase angles. 
Since there is no significant change in spectral slope between g=$13^\mathrm{o}$ and $30^\mathrm{o}$ we used the spectrum obtained at a phase angle of  $30^\mathrm{o}$ as reference point. 
Figure \ref{f:RatioedR2} shows the ratio plots (solid lines) between the spectra of the LL6 obtained at g=$60^\mathrm{o}$ and g=$30^\mathrm{o}$ (bottom), at g=$90^\mathrm{o}$ and 
g=$30^\mathrm{o}$ (middle) and at g=$120^\mathrm{o}$ and g=$30^\mathrm{o}$ (top panel). The ratio plots were modeled by fitting the same exponential curve used by \citet{2006Icar..184..327B}

\begin{equation}
{\rm{Ratio}} =  P_{R}(\lambda) = Aexp(C_{P}/\lambda)
\end{equation}

where we have just changed the name of the function, we call it the phase reddening function ($P_{R}$) and the parameters A and C$_{P}$ are equivalents to the parameters K and C$_{S}$ respectively, with the 
difference that C$_{P}$ would be a measure of the effect of the phase reddening. The best fit curves obtained using Eq. (13) are represented with dashed lines in Fig. \ref{f:RatioedR2}. Here we have used the LL6 
spectra as an example, but the same procedure can be used with the other samples. In Fig. \ref{f:RatioedR2} we have also included the resultant A and C$_{P}$ values and the coefficient of determination ($R^{2}$) 
of the fitted curves for each ratio plot. Similarly to the results obtained by \citet{2006Icar..184..327B} the ratio of the spectra show some traces of Bands I and II, which can affect the fitting in those spectral regions. 
Nevertheless, the $R^{2}$ for the fitted curves ranges between 0.88 and 0.96 and, in general, the results are comparable to those of  \citet{2006Icar..184..327B}. 
After comparing the C$_{P}$ values with the C$_{S}$ obtained by \citet{2006Icar..184..327B} for different samples, we found that the C$_{P}$ values are similar to the C$_{S}$ obtained for the H5 ordinary 
chondrite Epinal, which was irradiated with Ar$^{++}$ (60 keV) at three different ion fluences.     

\begin{figure*}[!ht]
\begin{center}
\psfig{file=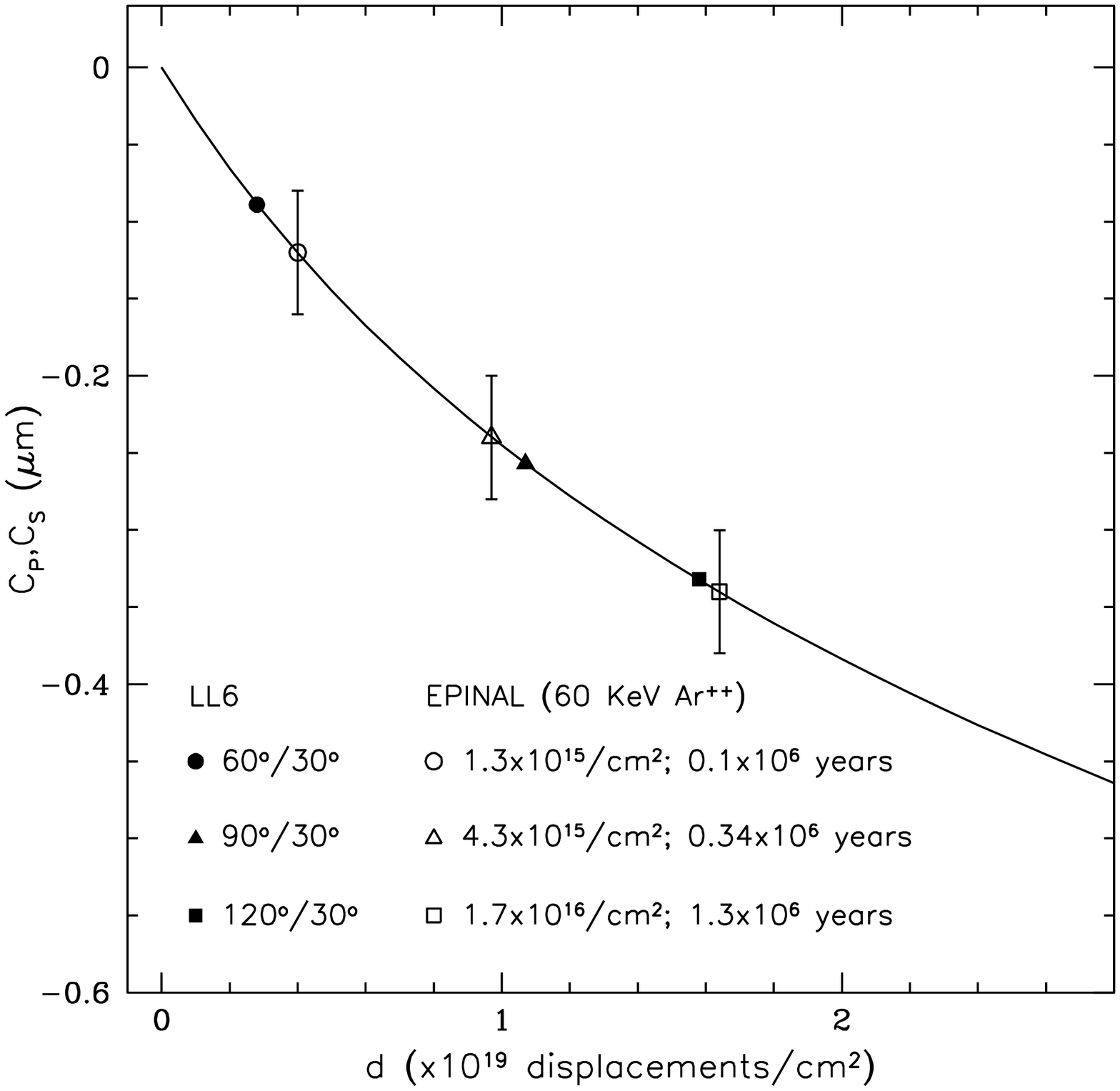,angle=0,height=11cm}
\caption{\label{f:C_d2} {\small The C$_{P}$ (LL6 ordinary chondrite) and C$_{S}$ (Epinal ordinary chondrite) parameters as a function of the number of displacements per cm$^{2}$ (damage parameter). The data 
corresponding to the Epinal meteorite were obtained from \citet{2006Icar..184..327B}. The LL6 and the Epinal data are plotted as filled and open symbols respectively. For the LL6 each symbol represents the value 
obtained from the ratioed spectra of different phase angles. The symbols corresponding to the Epinal meteorite represent different ions fluences. Approximate exposure times (at 1 AU) corresponding to 
each ion fluence are also shown. The solid line represents the experimental damage curve from \citet{2006Icar..184..327B}. The error bars of the LL6 data are on the order of the size of symbols.}}
\end{center}
\end{figure*}


\citet{2006Icar..184..327B} found that there is a strong correlation between the C$_{S}$ parameter and the number of displacements per cm$^{2}$ (damage parameter). From their experimental data they were able 
to create a damage curve that was fitted by

\begin{equation}
C_{S}=\alpha ln(\beta d + 1)
\end{equation}

where $d$ is the damage parameter, $\alpha = -0.33\pm 0.06$ $\mu$m and $\beta = (1.1\pm 0.5) \times 10^{-19}$ cm$^{2}$. Using this equation we calculated the number of displacements per cm$^{2}$ 
that would correspond to our C$_{P}$ values. It is important to point out that in this case the $d$ values calculated using the C$_{P}$ have no real physical meaning, since the C$_{P}$ parameter quantifies the effect of the 
phase reddening and not the space weathering, however their calculation is useful to illustrate how the phase reddening can resemble different degrees of space weathering. This can be seen in Figure \ref{f:C_d2} 
where we have plotted the C$_{P}$ (LL6 ordinary chondrite) and C$_{S}$ (Epinal ordinary chondrite) values versus the calculated damage parameter for each sample. The LL6 data are represented with filled symbols 
while the Epinal values are plotted as open symbols. These results indicate that an increase in phase angle in the range of $30^\mathrm{o}$ to $120^\mathrm{o}$ would be equivalent to irradiate the sample with high 
mass ions (Ar$^{++}$ 60 keV) with an ion fluence in the range of $\sim 1.3\times 10^{15}$ to $1.7\times 10^{16}$  ions cm$^{-2}$. These ion fluences are equivalent to exposure times of $\sim 0.1\times 10^{6}$ to 
$1.3\times 10^{6}$ years at about 1 AU from the Sun \citep[calculated as in][]{2005Icar..174...31S}.
Laboratory experiments \citep[e.g.,][]{2005Icar..174...31S} have shown that the timescale for the weathering of NEAs surfaces due to ion irradiation is on the order of $10^{4}$-$10^{6}$ years. Thus, observing NEAs at 
high phase angles could produce spectral slopes that resemble those exhibited by weathered surfaces.

For the second experiment we modeled the optical effects of the SMFe on the laboratory spectra by using Hapke's radiative transfer model \citep{1981JGR....86.3039H,1993tres.book.....H,2001JGR...10610039H}, 
and then we compared these spectra with those obtained at different phase angles. 

From \citet{2001JGR...10610039H} the bidirectional reflectance of a medium of isotropic scatterers ignoring the opposition effect is given 
by

\begin{equation}
r({\rm{i}},{\rm{e}},{\rm{g}})=\frac{w}{4\pi}\frac{\mu_{0}}{\mu_{0}+\mu}H(\gamma,\mu_{0})H(\gamma,\mu)
\end{equation}
   
where $i$, $e$ and g are the incidence, emission and phase angle respectively, $\mu_{0}$=cos(i), $\mu$=cos(e), {\it{w}} is the single scattering albedo, $\gamma=(1-w)^{1/2}$, and H($\gamma$, $\mu$) is an 
analytic approximation to the Ambartsumian-Chandrasekhar H functions. Since most reflectances are relative to a standard, \citet{2001JGR...10610039H} represented this relative reflectance as

\begin{equation}
\Gamma(\gamma)=\frac{1-\gamma^{2}}{(1+2\gamma \mu_{0})(1+2\gamma \mu)}
\end{equation}
   
from which $\gamma$ can be determined, and the single scattering albedo of the sample can be calculated as

\begin{equation}
w=1-\gamma^{2}
\end{equation}

The single scattering albedo can be also written as a function of the properties of the particles of the medium by

\begin{equation}
w=S_{e}+(1-S_{e})\frac{1-S_{i}}{1-S_{i}\Theta}\Theta
\end{equation}

where $S_{e}$ is the Fresnel reflection coefficient for externally incident light, $S_{i}$ is the Fresnel reflection coefficient for internally scattered light
 \citep[see][]{2001JGR...10610039H}, and $\Theta$ is the single-pass transmission of the particle. If there is no internal scattering, then 

\begin{equation}
\Theta=e^{-\alpha \langle D\rangle} 
\end{equation}

where $\langle D\rangle$ is the mean ray path length \citep{1993tres.book.....H} and $\alpha$ is the absorption coefficient given by

\begin{equation}
\alpha=\frac{4\pi nk}{\lambda}
\end{equation}

where {\it{n}} and {\it{k}} are the real and imaginary part of the refractive index respectively and $\lambda$ is the wavelength. 
From Eqs. (18) and (19) the absorption coefficient can be also calculated as

\begin{equation}
\alpha=\frac{1}{\langle D\rangle}ln\bigg[S_{i}+\frac{(1-S_{e})(1-S_{i}}{w-S_{e}}\bigg]
\end{equation}
 
To model the effects of the SMFe, the absorption coefficient of the host material ($\alpha_{h}$) is increased by adding to it the absorption coefficient of the SMFe ($\alpha_{Fe}$). Using the Maxwell-Garnett effective medium theory, \citet{2001JGR...10610039H} derived an expression to calculate $\alpha_{Fe}$,

\begin{equation}
\alpha_{Fe}=\frac{36\pi zf\rho_{h}}{\lambda \rho_{Fe}}
\end{equation}

where {\it{f}} is the mass fraction of the Fe particles, $\rho_{h}$ is the density of the host material, $\rho_{Fe}$ is the density of iron and  {\it{z}} is given by

\begin{equation}
z=\frac{{n^{3}}_{h}n_{Fe}k_{Fe}}{({n^{2}}_{Fe}-{k^{2}}_{Fe}+2{n^{2}}_{h})^{2}+(2n_{Fe}k_{Fe})^{2}}
\end{equation}

where $n_{h}$ and $n_{Fe}$ are the real part of the refractive indices of the host material and iron respectively, and $k_{Fe}$ is the imaginary part of the refractive index of iron. Thus, if we use Eq. (21) to calculate 
the absorption coefficient of the host material ($\alpha_{h}$), then the absorption coefficient of the material containing SMFe is given by

\begin{equation}
\alpha_{w}=\alpha_{h}+\alpha_{Fe}
\end{equation}
 
Looking at Eqs. (15) and (18)-(20) we can understand why phase reddening and space weathering are manifested in similar ways. Reflectance spectra are controlled by the single scattering albedo ({\it{w}}), 
which is a function of the absorption coefficient ($\alpha$). The absorption coefficient is a parameter that characterizes how deep into a material light of a particular wavelength can penetrate before being absorbed. 
As the phase angle increases, light is less able to escape from the medium, meaning that photons are more absorbed, resulting in a decrease of reflectance throughout the entire spectrum. However, because 
the absorption coefficient is inversely proportional to the wavelength (the shorter the wavelength the higher the absorption coefficient) the reflectance in the blue part of the spectrum will decrease faster than in the red 
part, producing the increase in spectral slope (reddening). Similarly, the addition of the SMFe (Eq. 24) will decrease the blue part of the spectrum more than the red part due to the stronger absorption at short 
wavelengths, increasing the spectral slope \citep{2001JGR...10610039H}. 

\begin{figure*}[!ht]
\begin{center}
\psfig{file=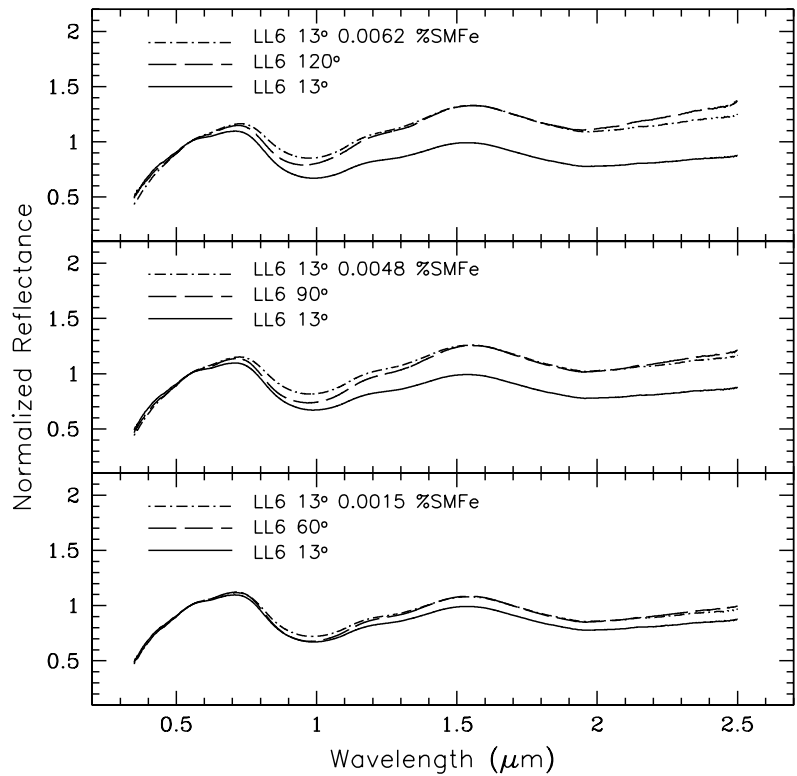,angle=0,height=11cm}
\caption{\label{f:SMFe2} {\small Reflectance spectra of the LL6 chondrite Dhurmsala after adding different amounts of SMFe (dot-short dashed lines) to the spectrum obtained at g=$13^\mathrm{o}$. For comparison the 
spectra of the LL6 obtained at g=$13^\mathrm{o}$ (solid lines) and g=$60^\mathrm{o}$, $90^\mathrm{o}$ and $120^\mathrm{o}$ (long dashed lines from the bottom to the top) have been included. All the spectra are 
normalized to unity at 0.55 $\mu$m.}}
\end{center}
\end{figure*}

For our experiment we used the spectrum of the LL6 ordinary chondrite obtained at a phase angle of $13^\mathrm{o}$ as reference point. Using these data we first calculated $\alpha_{h}$ with  Eqs. 16 to 21, where 
we have assumed $n_{h}=1.7$, which is a typical value for mafic minerals \citep{2001JGR...10610039H}.  Then using Eqs. 22 and 23 we estimated $\alpha_{Fe}$. For this calculation we have assumed 
$\rho_{h}=3.48$, which is the average grain density for LL chondrites \citep{2003M&PS...38.1161B} and $\rho_{Fe}=7.87$. The optical constants of iron, $n_{Fe}$ and $k_{Fe}$, were taken from 
\citet{1974PhRvB...9.5056J}. Since these values were measured only for $\lambda \sim 1.9$ $\mu$m, we did a polynomial fit to the data in order to extrapolate them to 2.5 $\mu$m. The $\alpha_{Fe}$ was 
calculated for different mass fractions of Fe and then the resulting values were added to the calculated $\alpha_{h}$ in order to obtain $\alpha_{w}$. These  $\alpha_{w}$ values were then inserted into equation 
(19), which combined with equation (18) allowed us to determine the new single scattering albedos. Finally, we calculated $\gamma$ from Eq. (17) and using Eq. (16) we obtained the "weathered spectra" for the 
different $\%$ of SMFe. In Figure \ref{f:SMFe2} we plot the reflectance spectra of the LL6 after adding different amounts of SMFe.  We found that the spectrum obtained at  ${\rm{g}}=60^\mathrm{o}$ shows a spectral 
slope comparable to the spectrum obtained at  ${\rm{g}}=13^\mathrm{o}$ after adding 0.0015 $\%$SMFe. For phase angles of 90$^\mathrm{o}$ and 120$^\mathrm{o}$ the spectral slopes are equivalent to adding 0.0048 
and 0.0062  $\%$SMFe respectively, to the spectrum obtained at  ${\rm{g}}=13^\mathrm{o}$. If we consider that an amount of $\sim$ 0.02$\%$ of SMFe is required to explain the red slopes of some S-complex asteroids 
\citep[e.g.,][]{2001JGR...10610039H,2004Icar..172..408R}, this means that the spectral slope for the highest phase angle (${\rm{g}}=120^\mathrm{o}$) would be equivalent to $\sim$ 30$\%$ of that SMFe.

 \section{Phase reddening and taxonomic classification}

The taxonomic classification of asteroids is based on shared observational parameters like spectral slope, color, albedo and band depth.
Most of the current classification systems are based on visible data \citep[e.g.,][]{1984PhDT.........3T,2002Icar..158..106B,2002Icar..158..146B}, in part because only during the last decade sufficient high-quality NIR 
spectral data became available to extend the classification to the near-infrared wavelengths. Two of the most common systems used to classify asteroids using VIS-NIR data are those introduced by 
\citet{1993Icar..106..573G} and more recently by \citet{2009Icar..202..160D}. \citet{1993Icar..106..573G} developed their classification system from the analysis of 39 asteroids classified as S-type by 
\citet{1984PhDT.........3T}. They divided the S-population into seven main compositional subgroups designated S(I)-S(VII). These subgroups range from pure olivine through olivine-pyroxene mixtures to pure 
pyroxene mixtures, and were derived on the basis of two band parameters, the Band I center and the BAR. The Bus-DeMeo taxonomy, on the other hand, is based on Principal Component Analysis and is 
comprised of 24 classes that include three major complexes (S-, C- and X-complex) and the end members O, Q, R, V, D, K, L, T. The S-complex is subdivided into S, Sa, Sq, Sr and Sv. Since both taxonomic systems 
use band depths as one of the primary criteria to classify the objects, multiple observations of the same asteroid obtained at different phase angles could lead to ambiguous classifications. In order to test the influence 
of phase reddening on the taxonomic classification, we have applied the two classification systems described above to each of the observed asteroids.                  

\begin{figure*}[!ht]
\begin{center}
\psfig{file=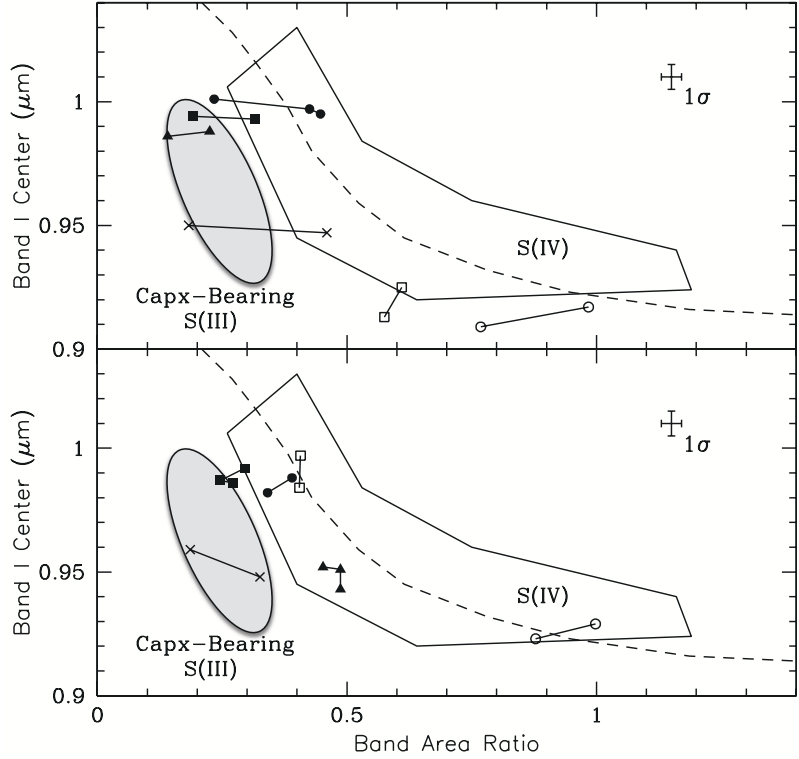,angle=0,height=11cm}
\caption{\label{f:BI_BAR6B} {\small Plot of the Band I center versus BAR for the NEAs. The polygonal region represents the mafic silicate components of ordinary chondrites and S(IV) asteroids 
\citep{1993Icar..106..573G}. The grey oval region represents the mineralogical zone corresponding to the calcic pyroxene-bearing where the S(III) subtypes are located \citep{1993Icar..106..573G}. 
The dashed curve indicates the location of the olivine-orthopyroxene mixing line \citep{1986JGR....9111641C}. The top panel shows the measured values for asteroids 1620 Geographos 
(filled circles), 1036 Ganymed (open circles), 1862 Apollo (filled triangles), 1980 Tezcatiploca (crosses), 11398 (open squares) and 6239 Minos (filled squares). The bottom panel shows the measured values for 
asteroids 4954 Eric (open circles), 66146 (filled squares), 1627 Ivar (crosses), 35107 (open squares), 25143 Itokawa (filled circles) and 4179 Toutatis (filled triangles). Measured values from multiple 
observations of individual asteroids are connected by lines. The average 1-$\sigma$ error bars are shown in the upper right corner.}}
\end{center}
\end{figure*}

\begin{figure*}[!ht]
\begin{center}
\psfig{file=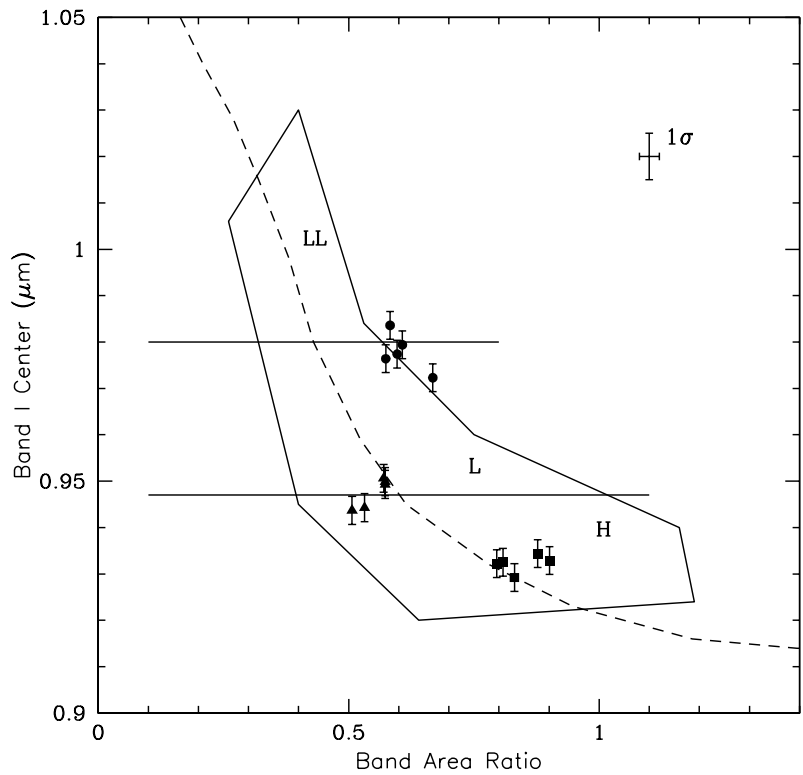,angle=0,height=11cm}
\caption{\label{f:b1_BAROC3} {\small Plot of the Band I center versus BAR for the LL6 (circles), L6 (triangles) and H6 (squares) ordinary chondrites. The polygonal region represents the mafic silicate components of 
ordinary chondrites and S(IV) asteroids \citep{1993Icar..106..573G}. The dashed curve indicates the location of the olivine-orthopyroxene mixing line \citep{1986JGR....9111641C}. The horizontal lines represent 
the approximate boundaries that separate the three types of ordinary chondrites found by \citet{2010Icar..208..789D}. 
The error bar in the x-axis is of the order of the size of symbols. For comparison purpose we have included the average 1-$\sigma$ error bars (upper right corner) 
associated with the Band I center and BAR value measured for asteroids.}}
\end{center}
\end{figure*}

Figure \ref{f:BI_BAR6B} shows the measured Band I center versus BAR for the NEAs. 
Within the uncertainties all the NEAs studied are classified either as S(III) or S(IV). There are four cases (1620 Geographos, 1627 Ivar, 1980 Tezcatlipoca and 1036 Ganymed) that show large variations of the BAR from 
one phase angle to another, but this is likely attributed to noisy spectra rather than phase reddening. The rest of the asteroids show variations in Band I centers and BAR that are on the order of the 
error bars. The results obtained from the laboratory samples (Fig. \ref{f:b1_BAROC3}) indicate that the largest variation in the BAR ($\sim$ 0.1) is larger than the typical uncertainty associated with the BAR 
values measured from asteroids (represented by the 1-$\sigma$ error bars in the upper corner of this figure). This means that in certain cases phase reddening could contribute to an ambiguous classification, 
particularly if the measured band parameters of the object are located close to the boundaries that define each class.

\begin{figure*}[!ht]
\begin{center}
\psfig{file=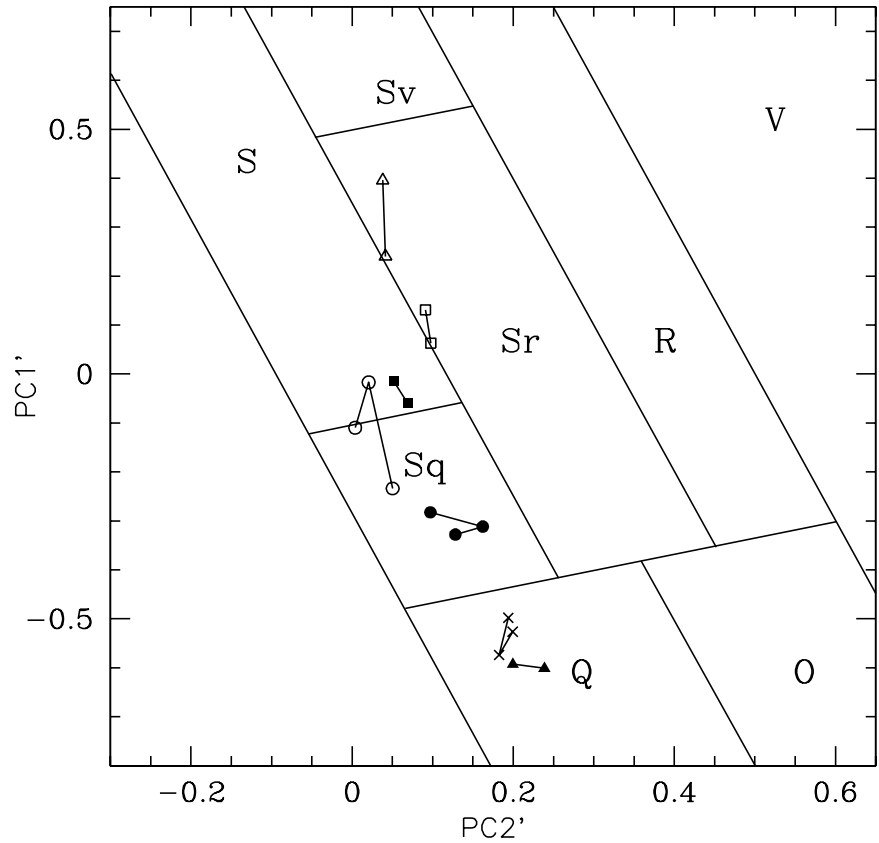,angle=0,height=12cm}
\caption{\label{f:pc1_pc22} {\small Calculated PC values for some of the NEAs studied represented in a PC1' versus PC2' diagram from \citet{2009Icar..202..160D}. The letters indicate the different classes within 
the S-complex plus Q-, O-, R- and V-types. The asteroids whose calculated PC values have been plotted are: (1036) Ganymed (open triangles), (1620) Geographos (open circles), (1627) Ivar (filled squares), 
(4179) Toutatis (filled circles), (1862) Apollo (filled triangles), 11398 (open squares) and 66146 (crosses).}}
\end{center}
\end{figure*}

The classification into the Bus-DeMeo system was done using the online taxonomy calculator (http://smass.mit.edu/busdemeoclass.html). The class assigned to each spectrum and the calculated principal 
components PC1' and PC2' are given in Table \ref{t:Table3}. Only in three cases asteroids were ambiguously classified. Asteroid (1036) Ganymed was classified as S and Sr, (1620) Geographos was classified as 
S and Sq and (4954) Eric was classified as Sw and Sr. The letter "w" was introduced by \citet{2009Icar..202..160D} to indicate an object exhibiting a high slope, but does not represent a different class, 
thus an object designated as "Sw" is an asteroid classified as S-type that exhibit a higher slope than the typical S-types. This distinction is based on an arbitrary cutoff at slope = 0.25 \citep{2009Icar..202..160D}.          
Table \ref{t:Table3} shows that the notation of "w" was added to many asteroids from one phase angle to another. Certainly the phase reddening most play a role on this variation in spectral slope, however 
as it was stated earlier this is a very sensitive parameter that can be affected by other factors. What is important to point out is that in the Bus-DeMeo system the spectral slope is removed prior to the classification.
For those objects which were given two classifications it is likely that the phase reddening was not the only contribution, since some of the spectra show more scatter than others. However from our 
analysis we estimate an average variation of about 0.04 and 0.03 for the PC1' and PC2' respectively, that could be attributed to phase reddening. These results suggest that phase reddening could lead to an 
ambiguous classification but only if the calculated PC values are close to the lines that separate each class. This can be seen in Figure \ref{f:pc1_pc22} where we have plotted the PC values in a PC1' versus PC2' 
diagram from \citet{2009Icar..202..160D}. For clarity only the calculated PC values for some of the asteroids have been plotted. These values are represented by different symbols, which are connected by lines to 
indicate multiple values for individual asteroids.


\clearpage

\section{Conclusions}

The analysis of VIS-NIR spectra of 12 NEAs observed at different phase angles has revealed an increase of band depths with increasing phase angles in the range of $2^\mathrm{o}<{\rm{g}}< 70^\mathrm{o}$ for 
Band I and  $2^\mathrm{o}<{\rm{g}}< 55^\mathrm{o}$ for Band II. Our analysis showed that, on average, Band I and Band II depths will increase 0.66\% and 0.93\% respectively, for every 10$^\mathrm{o}$ 
increase in phase angle. From the available data we have derived equations that can be used to correct the effects of phase reddening in the band depths.
Small variations in band centers and BAR values with increasing phase angle were also found. Similar trends were observed in the laboratory spectra of three different types of ordinary chondrites. In addition, an 
increase in the spectral slope with increasing phase angles was found for the ordinary chondrites. This increase in spectral slope is more significant for phase angles higher than $30^\mathrm{o}$. Of the three types 
of ordinary chondrites the (olivine-rich) LL6 is the most affected by phase reddening, showing the largest variations in spectral slopes and band depths with increasing phase angles. These variations in spectral band 
parameters seems to have no significant impact on the mineralogical analysis though. We have also found that the increase in spectral slope caused by phase reddening can mimic the effect of space 
weathering. In particular, an increase in phase angle in the range of $30^\mathrm{o}$ to $120^\mathrm{o}$ will produce a reddening of the reflectance spectra equivalent to exposure times of $\sim 0.1\times 10^{6}$ 
to $1.3\times 10^{6}$ years at about 1 AU from the Sun. Furthermore, the increase in spectral slope due to phase reddening is comparable to the effects caused by the addition of various amounts of SMFe. These 
results imply that phase reddening should be considered when studying space weathering effects on spectral data. Regarding to the taxonomic classification, we found that phase reddening can lead to an 
ambiguous classification, but only if the measured spectral band parameters and/or principal components (depending on the taxonomic system) are located close to the boundaries that define each class.

\clearpage

\begin{table}[!ht]
\caption{\label{t:Table1} {\small NEAs observational circumstances. The columns in this table are: object number and designation, UT date of observation, the phase angle (g) and the heliocentric 
distance at the time of observation.}}
\begin{center}
\begin{tabular}{|c|c|c|c|}
 \hline
Object&UT Date&g ($^\mathrm{o}$)&Heliocentric Distance (AU)\\ \hline
1036 Ganymed&09-Mar-05$^{b}$&5.9&4.09 \\ 
1036 Ganymed&01-Jun-06$^{c}$&15.2&3.01 \\ 
1620 Geographos&10-Mar-08$^{c}$&38.3&1.09 \\ 
1620 Geographos&27-Feb-08$^{a}$&12.0&1.16 \\ 
1620 Geographos&29-Jan-01$^{b}$&16.7&1.39 \\
1627 Ivar&02-Oct-08$^{c}$&31.0&1.58  \\  
1627 Ivar&03-Dec-08$^{c}$&16.0&1.90 \\ 
1862 Apollo&13-Nov-05$^{c}$&51.7&1.05 \\ 
1862 Apollo&22-Nov-05$^{c}$&15.4&1.20 \\
1980 Tezcatlipoca&25-Oct-06$^{c}$&54.6&1.17 \\ 
1980 Tezcatlipoca&05-Jan-07$^{a}$&25.0&1.48 \\ 
4179 Toutatis&15-Sep-04$^{c}$&27.0&1.09 \\ 
4179 Toutatis&02-Oct-08$^{c}$&68.0&1.06 \\ 
4179 Toutatis&13-Aug-08$^{a}$&6.1&1.45 \\ 
4954 Eric&20-Jul-07$^{c}$&28.3&1.67 \\ 
4954 Eric&26-Nov-07$^{a}$&62.0&1.10 \\ 
6239 Minos&26-Jan-04$^{b}$&47.3&1.03 \\ 
6239 Minos&06-Sep-10$^{c}$&2.4&1.20 \\
11398&10-Mar-08$^{c}$&33.0&1.16 \\ 
11398&28-Feb-08$^{a}$&19.0&1.21 \\  
25143 Itokawa&12-Mar-01$^{d}$&25.6&1.07 \\ 
25143 Itokawa&28-Mar-01$^{e}$&69.0&1.01 \\ 
35107&26-Jul-08$^{a}$&87.0&1.08 \\
35107&27-Dec-02$^{b}$&45.0&1.20 \\ 
66146&02-Oct-08$^{c}$&70.3&1.06 \\
66146&13-Oct-10$^{c}$&32.1&1.15 \\
66146&06-Sep-10$^{c}$&55.3&1.16 \\  \hline
\end{tabular}
\end{center}
{\small $^{a}$ Data from \citet{Reddy09}}

{\small$^{b}$ Data from \citet{2009Icar..202..160D}}

{\small$^{c}$ Data from NEOSR, http://smass.mit.edu/minus.html}

{\small$^{d}$ Data from \citet{2007M&PS...42.2165A}}

{\small$^{e}$ Data from \citet{2001M&PS...36.1167B}}

\end{table}

\clearpage

\begin{table}[!ht]
\caption{\label{t:Table2} {\small NEAs spectral band parameters and their errors. The columns in this table correspond to: object number and designation, the phase angle (g), the average 
surface temperature (T),  the Band I center (BI$\pm$0.01), the Band I depth (BI$_{\rm{dep}}\pm$0.3), the Band II center (BII$\pm$0.03), the temperature corrected Band II center ($\Delta$BII$\pm$0.03), the Band II 
depth (BII$_{\rm{dep}}\pm$0.5), the temperature corrected Band II depth ($\Delta$BII$_{\rm{dep}}\pm$0.5), the band area ratio (BAR$\pm$0.04) and the temperature corrected band area ratio ($\Delta$BAR$\pm$0.04).}}
\begin{center}\small
\begin{tabular}{|c|c|c|c|c|c|c|c|c|c|c|}
\hline
Object&g&T&BI&BI$_{\rm{dep}}$&BII&$\Delta$BII&BII$_{\rm{dep}}$&$\Delta$BII$_{\rm{dep}}$&BAR&$\Delta$BAR \\ 
 &($^\mathrm{o}$)&(K)&($\mu$m)&(\%)&($\mu$m)&($\mu$m)&(\%)&(\%)& &  \\ \hline
1036 Ganymed&5.9&129.6&0.909&18.79&1.867&1.901&8.05&5.49&0.901&0.768 \\
1036 Ganymed&15.2&150.9&0.917&18.22&1.883&1.913&9.29&7.05&1.101&0.984 \\
1620 Geographos&12.0&240.4&0.995&16.34&1.970&1.982&5.68&4.79&0.497&0.447 \\
1620 Geographos&16.7&219.6&0.997&16.77&1.967&1.984&6.56&5.35&0.490&0.425 \\
1620 Geographos&38.3&247.7&1.001&18.03&2.013&2.023&4.19&3.40&0.278&0.234 \\
1627 Ivar&16.0&198.9&0.948&19.14&1.955&1.975&6.26&4.74&0.407&0.326 \\
1627 Ivar&31.0&218.4&0.959&20.28&1.964&1.980&5.04&3.82&0.253&0.186 \\
1862 Apollo&15.4&243.3&0.986&25.00&1.997&2.008&4.81&3.96&0.188&0.141 \\
1862 Apollo&51.7&260.0&0.988&27.54&1.989&1.997&6.53&5.93&0.260&0.225 \\
1980 Tezcatlipoca&25.0&218.6&0.947&16.05&1.932&1.949&6.62&5.40&0.526&0.460 \\
1980 Tezcatlipoca&54.6&246.2&0.950&20.45&1.946&1.956&4.37&3.56&0.228&0.183 \\
4179 Toutatis&6.1&222.3&0.943&14.91&1.971&1.987&4.89&3.72&0.551&0.487 \\
4179 Toutatis&27.0&256.9&0.951&17.44&1.973&1.981&7.29&6.64&0.524&0.487 \\
4179 Toutatis&68.0&260.5&0.952&18.04&1.962&1.970&6.90&6.31&0.487&0.452 \\
4954 Eric&28.3&207.1&0.929&18.74&1.925&1.944&9.78&8.39&1.073&0.998 \\
4954 Eric&62.0&255.2&0.923&16.69&1.946&1.955&8.81&8.14&0.917&0.878 \\
6239 Minos&2.4&244.1&0.994&16.56&2.030&2.041&4.03&3.19&0.239&0.192 \\
6239 Minos&47.3&263.2&0.993&19.02&2.010&2.017&6.10&5.55&0.348&0.315 \\
11398&19.0&243.3&0.913&19.04&1.935&1.946&6.89&6.04&0.622&0.575 \\
11398&33.0&248.3&0.925&19.81&1.945&1.955&8.33&7.55&0.653&0.610 \\
25143 Itokawa&25.6&229.2&0.988&17.02&2.017&2.030&6.07&5.01&0.448&0.390 \\
25143 Itokawa&69.0&235.2&0.982&19.85&2.017&2.030&5.45&4.48&0.395&0.341 \\
35107&45.0&244.1&0.984&19.57&2.019&2.030&7.40&6.56&0.452&0.405 \\
35107&87.0&257.6&0.997&18.90&1.996&2.004&6.92&6.28&0.444&0.407 \\
66146&32.1&250.1&0.992&22.17&1.983&1.993&6.20&5.45&0.339&0.296 \\
66146&55.3&249.1&0.987&22.99&1.984&1.994&6.35&5.59&0.290&0.246 \\
66146&70.3&260.2&0.986&22.57&1.992&2.000&5.75&5.15&0.306&0.272 \\ \hline
\end{tabular}
\end{center}
\end{table}

\clearpage

\begin{table}[!ht]
\caption{\label{t:Table3} {\small Olivine abundance with its error and taxonomy for the NEAs.  The columns in this table are: object number and designation, the phase angle (g), 
the olivine-pyroxene abundance ratio (ol/(ol+px)$\pm$0.03), the temperature corrected olivine-pyroxene ratio ($\Delta$ol/(ol+px)$\pm$0.03), taxonomic classification (Bus-DeMeo) and principal components PC1' 
and PC2' for each asteroid.}}
\begin{center}\small
\begin{tabular}{|c|c|c|c|c|c|c|}
\hline
Object&g ($^\mathrm{o}$)&ol/(ol+px)&$\Delta$ol/(ol+px)&Taxonomy&PC1'&PC2' \\ \hline 
1036 Ganymed&5.9&0.51&0.54&S&0.2400&0.0416 \\
1036 Ganymed&15.2&0.46&0.49&Sr&0.3957&0.0381 \\
1620 Geographos&12.0&0.61&0.62&Sq&-0.1096&0.0038 \\
1620 Geographos&16.7&0.61&0.63&S&-0.0169&0.0206 \\
1620 Geographos&38.3&0.66&0.67&Sqw&-0.2337&0.0503 \\
1627 Ivar&16.0&0.63&0.65&S&-0.0151&0.0517 \\
1627 Ivar&31.0&0.67&0.68&Sw&-0.0591&0.0689 \\
1862 Apollo&15.4&0.68&0.69&Q&-0.5926&0.1996 \\
1862 Apollo&51.7&0.66&0.67&Q&-0.6011&0.2387 \\
1980 Tezcatlipoca&25.0&0.60&0.62&Sw&0.1132&0.0508 \\
1980 Tezcatlipoca&54.6&0.67&0.68&Sw&-0.0238&0.0588 \\
4179 Toutatis&6.1&0.59&0.61&Sqw&-0.2825&0.0972 \\
4179 Toutatis&27.0&0.60&0.61&Sq&-0.3115&0.1622 \\
4179 Toutatis&68.0&0.61&0.62&Sq&-0.3278&0.1282 \\
4954 Eric&28.3&0.47&0.49&Sw&0.1552&0.0433 \\
4954 Eric&62.0&0.51&0.52&Sr&0.4719&-0.0369 \\
6239 Minos&2.4&0.67&0.68&Sq&-0.2907&0.0557 \\
6239 Minos&47.3&0.64&0.65&Sqw&-0.3090&0.1231 \\
11398&19.0&0.58&0.59&Sr&0.0631&0.0978 \\
11398&33.0&0.57&0.58&Sr&0.1312&0.0912 \\
25143 Itokawa&25.6&0.62&0.63&Sq&-0.3425&0.0946 \\
25143 Itokawa&69.0&0.63&0.65&Sqw&-0.3283&0.1579 \\
35107&45.0&0.62&0.63&Sq&-0.2848&0.1281 \\
35107&87.0&0.62&0.63&Sq&-0.3077&0.1182 \\
66146&32.1&0.65&0.66&Q&-0.5262&0.1994 \\
66146&55.3&0.66&0.67&Q&-0.5737&0.1824 \\
66146&70.3&0.65&0.66&Q&-0.4978&0.1938  \\ \hline
\end{tabular}
\end{center}
\end{table}

\clearpage

\begin{table}[!ht]

\begin{center}\small
\caption{\label{t:Table4} {\small Spectral band parameters of the ordinary chondrites. The columns in this table correspond to: sample type, the incidence angle (i), the 
emission angle (e), the phase angle (g), the spectral slope $\pm$0.004, the Band I center (BI$\pm$0.003), the Band I depth (BI$_{\rm{dep}}\pm$0.1), the Band II center (BII$\pm$0.005), the Band II depth 
(BII$_{\rm{dep}}\pm$0.2), the band area ratio (BAR$\pm$0.01) and the olivine-pyroxene abundance ratio (ol/(ol+px)$\pm$0.03).}} 
\begin{tabular}{|c|c|c|c|c|c|c|c|c|c|c|}
\hline
Sample&i&e&g&Slope&BI&BI$_{\rm{dep}}$&BII&BII$_{\rm{dep}}$&BAR&ol/(ol+px) \\ 
 &($^\mathrm{o}$)&($^\mathrm{o}$)&($^\mathrm{o}$)&($\mu$$m^{-1}$)&($\mu$m)&(\%)&($\mu$m)&(\%)& &  \\ \hline
LL6&13&0&13&-0.127&0.984&36.81&1.963&17.27&0.58&0.59 \\
LL6&30&0&30&-0.094&0.982&38.41&1.959&18.35&0.58&0.59 \\
LL6&0&30&30&-0.083&0.978&38.47&1.961&18.73&0.61&0.58 \\
LL6&-30&60&30&-0.179&0.978&36.69&1.964&18.36&0.63&0.58 \\
LL6&30,0,-30&0,30,60&30&-0.119&0.979&37.86&1.961&18.48&0.61&0.58 \\
LL6&60&0&60&0.002&0.981&40.28&1.957&18.55&0.55&0.59 \\
LL6&0&60&60&-0.097&0.969&37.30&1.965&18.89&0.68&0.56 \\
LL6&30&30&60&-0.043&0.982&39.61&1.960&18.22&0.56& 0.59 \\
LL6&60,0,30&0,60,30&60&-0.046&0.977&39.06&1.961&18.55&0.60&0.58 \\
LL6&60&30&90&0.099&0.973&37.91&1.963&18.75&0.62&0.58 \\
LL6&30&60&90& 0.184&0.980&36.84&1.961&16.46&0.53&0.60 \\
LL6&60,30&30,60&90&0.141&0.976&37.38&1.962&17.61&0.57&0.59 \\
LL6&60&60&120&0.212&0.972&34.18&1.959&17.50&0.67&0.57 \\ 
L6&13&0&13&-0.102&0.949&42.37&1.944&18.70&0.57&0.59 \\
L6&30&0&30&-0.090&0.951&42.71&1.943&18.92&0.56&0.59 \\
L6&0&30&30&-0.089&0.950&42.39&1.945&19.03&0.58&0.59 \\
L6&-30&60&30&-0.192&0.950&41.58&1.947&18.00&0.57&0.59 \\
L6&30,0,-30&0,30,60&30&-0.124&0.951&42.23&1.945&18.65&0.57&0.59 \\
L6&60&0&60&-0.084&0.954&43.16&1.945&18.21&0.51&0.60 \\
L6&0&60&60&-0.085&0.949&43.36&1.948&18.73&0.59&0.59 \\
L6&30&30&60&0.004&0.947&44.68&1.939&20.86&0.62&0.58 \\
L6&60,0,30&0,60,30&60&-0.055&0.950&43.73&1.944&19.26&0.57&0.59 \\
L6&60&30&90&0.097&0.945&42.53&1.941&17.21&0.51&0.61 \\
L6&30&60&90&0.177&0.944&41.69&1.948&17.24&0.56&0.59 \\
L6&60,30&30,60&90&0.137&0.944&42.11&1.945&17.23&0.54&0.60 \\
L6&60&60&120&0.191&0.944&39.84&1.943&16.15&0.51&0.61 \\
H6&13&0&13&-0.141&0.934&35.45&1.930&18.41&0.88&0.52 \\
H6&30&0&30&-0.151&0.935&36.20&1.929&19.12&0.89&0.51 \\
H6&0&30&30&-0.115&0.933&35.33&1.934&18.33&0.88&0.52 \\
H6&-30&60&30&-0.177&0.931&34.80&1.931&18.26&0.93&0.50 \\
H6&30,0,-30&0,30,60&30&-0.148&0.933&35.44&1.932&18.57&0.90&0.51 \\
H6&60&0&60&-0.149&0.935&36.48&1.926&18.91&0.83&0.53 \\
H6&0&60&60&-0.096&0.929&34.54&1.935&16.93&0.84&0.52 \\
H6&30&30&60&-0.096&0.933&36.93&1.931&17.85&0.76&0.55 \\
H6&60,0,30&0,60,30&60&-0.114&0.933&35.98&1.931&17.90&0.81&0.53 \\
H6&60&30&90&-0.058&0.932&34.77&1.928&17.65 &0.83&0.53 \\
H6&30&60&90&-0.012&0.926&31.69&1.934&14.77&0.84&0.53 \\
H6&60,30&30,60&90&-0.035&0.929&33.23&1.931&16.21&0.83&0.53 \\
H6&60&60&120&0.020&0.932&31.77&1.932&15.55&0.80&0.54 \\ 
\hline

\end{tabular}
\end{center}
\end{table}

\clearpage


{\bf{Acknowledgements}}

\

The authors would like to thank, Francesca DeMeo, John Hinrichs and Paul Lucey for providing us with data for this research. We also thank to Michael Gaffey for the meteorite samples and 
Stefan Schr$\ddot{o}$der and Silvia Protopapa for the fruitful discussions related to this work. We would also like to thank Tasha Dunn and Jian-Yang Li for their reviews, which helped to improve the manuscript. 
J. A. Sanchez acknowledges a PhD fellowship of the International Max Planck Research School on 
Physical Processes in the Solar System and Beyond. Vishnu Reddy's research is supported by NASA NEOO Program Grant NNX07AL29G, and NASA Planetary Geology and Geophysics Grant NNX07AP73G. 
EAC thanks the Canada Foundation for Innovation, the Manitoba Research Innovations Fund, and the Canadian Space Agency for their support of the establishment of the University of Winnipeg Planetary 
Spectrophotometer Facility, and NSERC for a Discovery grant, and the University of Winnipeg for various internal grants, to support this project.

\clearpage

 \appendix
\section{Asteroid spectra}
\label{}
This appendix contains the spectra of all NEAs analyzed in this study. The numerical designation and the date of the observation (YYMMDD) for each asteroid are given.

\begin{figure*}[!ht]
\begin{center}
\psfig{file=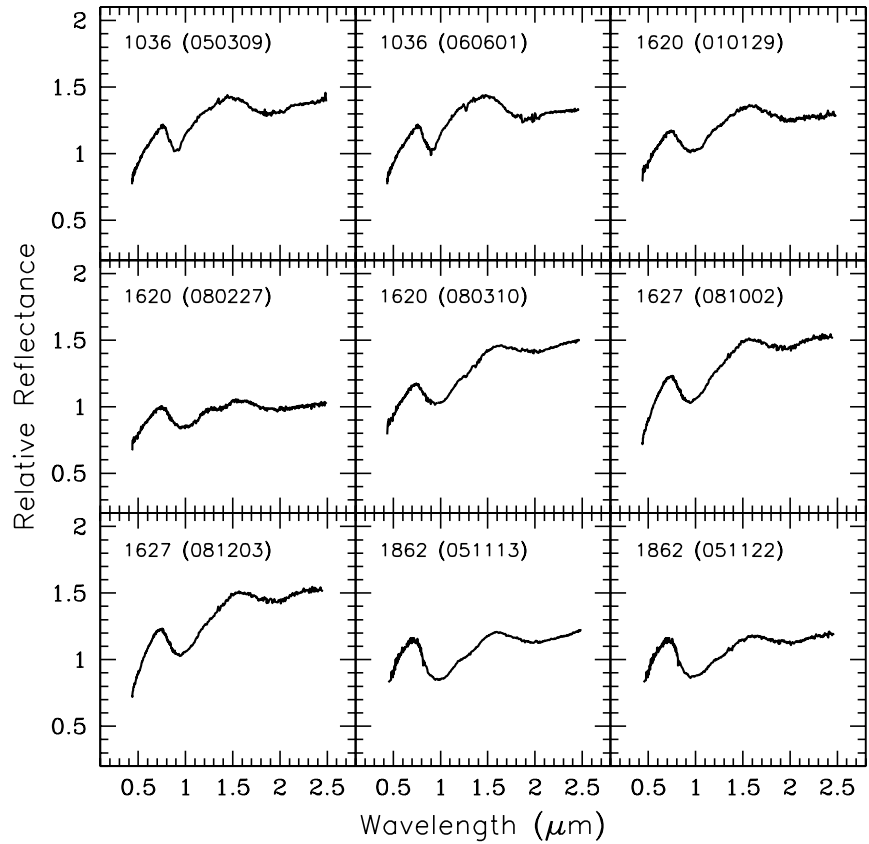,angle=0,height=12cm}
\caption{\label{f:asteroids} {\small Reflectance spectra of NEAs analyzed in this study.}}
\end{center}
\end{figure*}

\begin{figure*}[!ht]
\begin{center}
\psfig{file=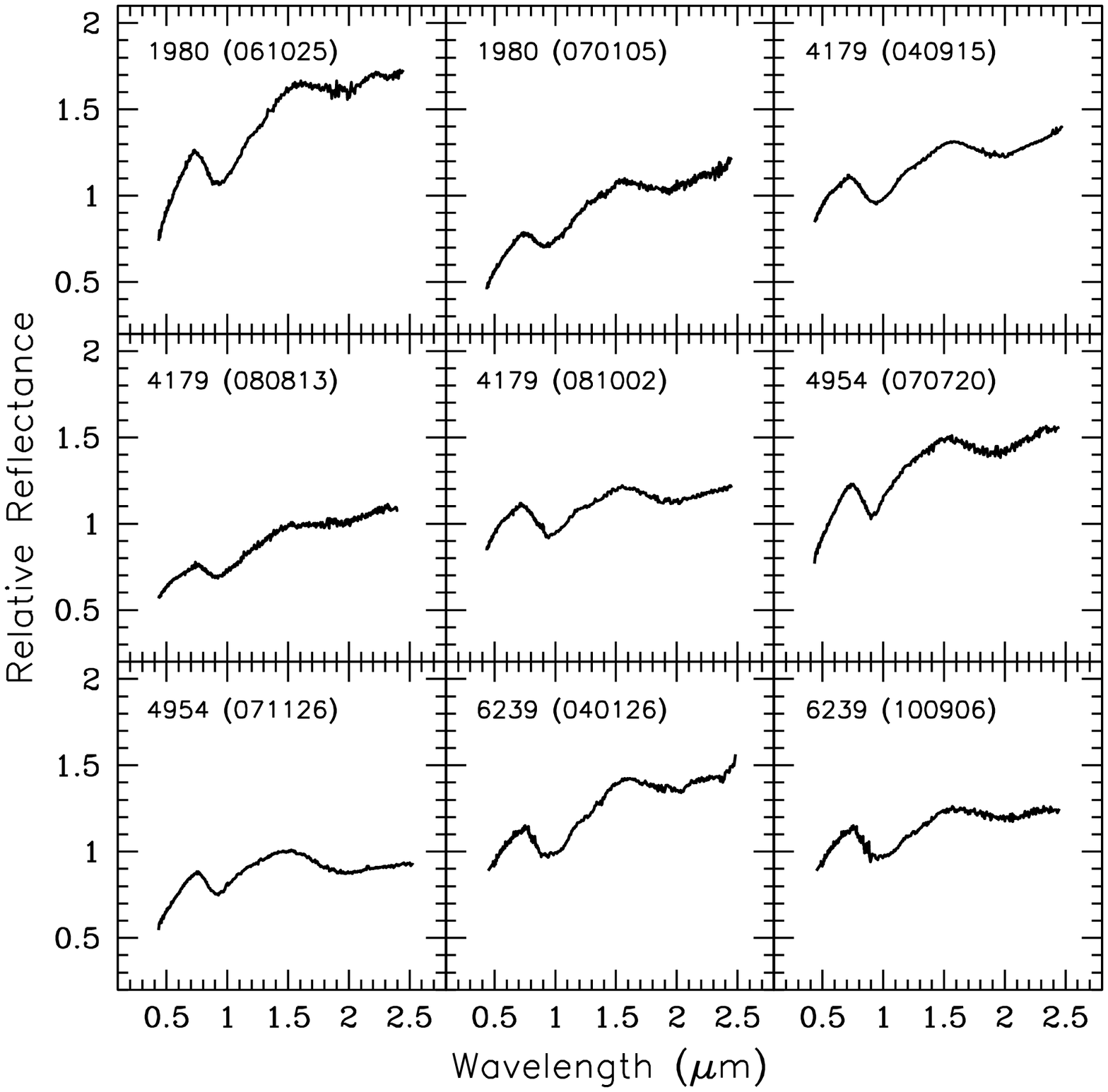,angle=0,height=12cm}
\caption{\label{f:asteroids2} {\small Reflectance spectra of NEAs analyzed in this study.}}
\end{center}
\end{figure*}

\begin{figure*}[!ht]
\begin{center}
\psfig{file=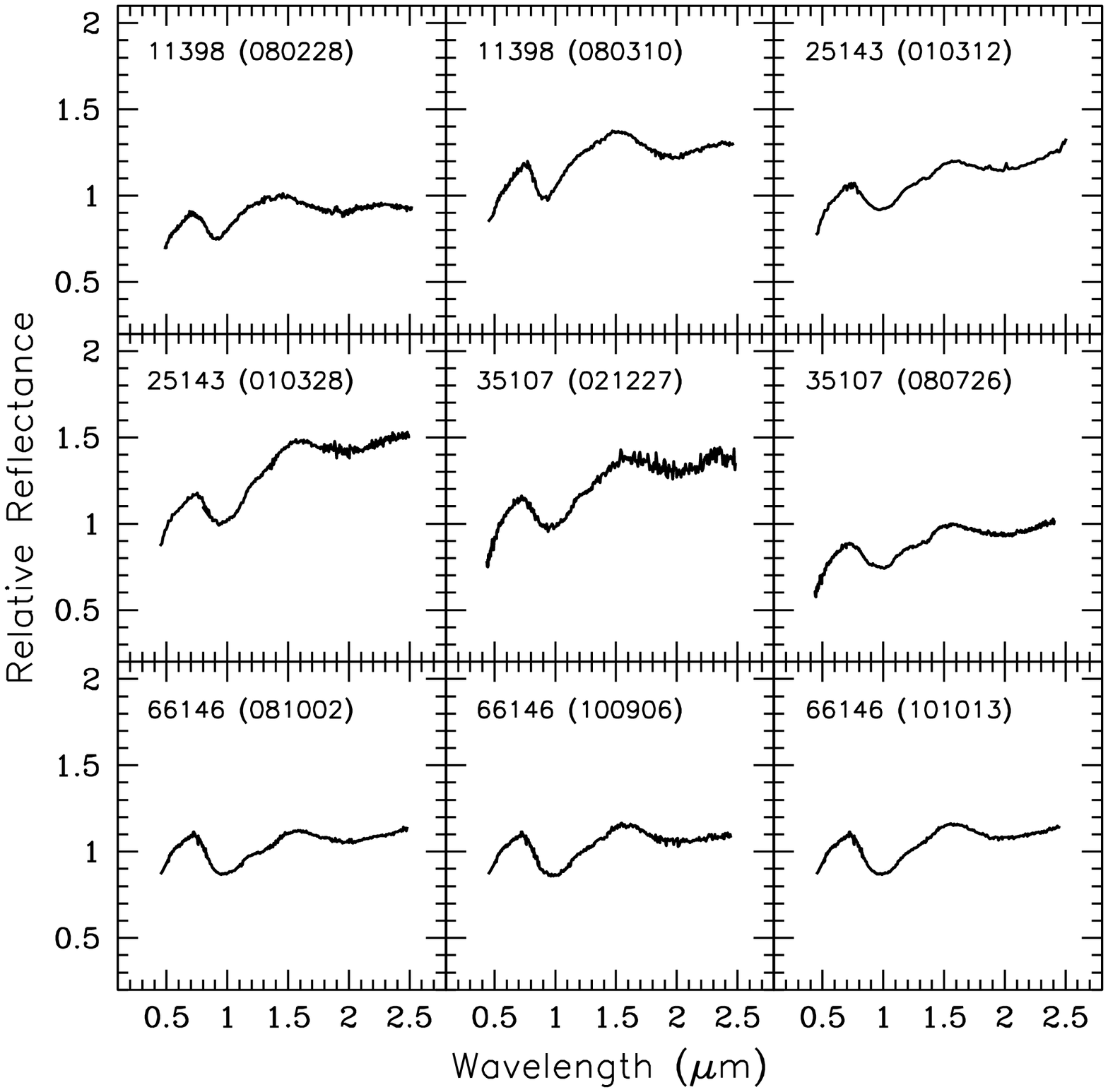,angle=0,height=12cm}
\caption{\label{f:asteroids3} {\small Reflectance spectra of NEAs analyzed in this study.}}
\end{center}
\end{figure*}

\clearpage

\bibliographystyle{plainnat}
\bibliography{references}







\end{document}